\documentclass{emulateapj}

\newcommand{\bl}[1]{\mbox{\boldmath$ #1 $}}

\shorttitle{Propagating star formation in Arp 10}
\shortauthors{Bizyaev et al.}

\begin{document}
\title{Propagating Star Formation in the Collisional Ring Galaxy Arp 10}

\author{Bizyaev D.V.\altaffilmark{1,2}, Moiseev A.V.\altaffilmark{3},
and Vorobyov E.I.\altaffilmark{4,5}}
\email{dmbiz@noao.edu, moisav@sao.ru, vorobyov@astro.uwo.ca}

\altaffiltext{1}{National Optical Astronomy Observatory, Tucson, AZ, 85719}
\altaffiltext{2}{Sternberg Astronomical Institute, Moscow, 119899, Russia} 
\altaffiltext{3}{Special Astrophysical Observatory, Nizhniy Arkhyz,
Karachaevo-Cherkessiya, Russia}
\altaffiltext{4}{CITA National Fellow, The University of Western
Ontario, London, Canada}
\altaffiltext{5}{Institute of Physics, Stachki 194, Rostov-on-Don, Russia}

\begin{abstract}

Propagating star formation in a collisional ring galaxy Arp~10 is
investigated by a complex approach, which includes the broad- and
narrow-band photometry, long-slit spectroscopy, and scanning
Fabry-Perot spectroscopy. The ionized gas velocity field obtained
with best spatial resolution to date indicates a
non-isotropic expansion of the outer ring with a maximum velocity 
110~km~s$^{-1}$. Strong vertical and non-circular motions
are also seen in the vicinity of the inner ring. Our kinematic data
suggest that Arp~10 has a small inclination $i=22\degr$ and high
total (luminous plus dark matter) mass of about $10^{12}~
M_{\odot}$ within a 50~kpc radius. The internal
extinction in HII regions and extinction-corrected  H$\alpha$ fluxes 
are estimated from the emission lines. The abundance of oxygen
12 + log(O/H) in both star-forming rings is about 8.6.  
The analysis of spectral indices provides an estimate on the 
propagation velocities of both rings and metallicity of
the pre-collision stellar population. 
A small "knot" near the nucleus of Arp~10, which was
previously suspected as a possible candidate for collision, 
is now unambiguously identified as the "intruder".  The
intruder spectrum suggests that it was a spiral galaxy before
the collision and its present mass amounts to at least one-fourth of
the total mass of Arp~10. A high value of the intruder velocity relative 
to Arp~10 ($\approx 480$~km~s$^{-1}$) suggests that the intruder is either
gravitationally unbound or marginally bound to the main galaxy.
We use a simplified two-dimensional hydrodynamic modeling of galaxy 
collisions to test a collisional origin of Arp~10. 
We confirm that the sizes of the inner and outer rings,  
maximum expansion velocity of the outer ring, and radial profile of the 
gas circular velocity can be reproduced by a near-central
collision with the intruder galaxy, which occurred 
approximately 85~Myr ago. We acknowledge that an apparent
crescent-shaped distribution of H$\alpha$ emission in the outer
ring is caused by a star formation threshold in the gas disk of
Arp~10.

\end{abstract}


\section{Introduction} \label{s0}

Collisional ring galaxies \citep{VV76} give us a bright example of 
star formation induced by radially expanding density waves, which are triggered 
by a near-axial passage of 
a companion galaxy through the disk of a target galaxy \citep{Lynds76,Higdon95,CA96}.
The expanding density waves in the gas disk of a target galaxy leave behind 
stellar populations of progressively older age.
This allows us to study the sequence of large-scale star formation
events by considering adjacent locations along the galactic radius.
A sequential triggering of star formation by expanding density waves 
makes collisional ring galaxies a unique laboratory for studying young 
stellar populations. 

The best studied collisional ring galaxy to date is the Cartwheel. 
Star formation in the Cartwheel is concentrated in the two rings, which 
reveal the presence of expanding density waves \citep{Higdon95,Higdon96,Appleton96}.  
The non-stationary and propagating nature of star formation is manifested by 
the current radial expansion of the rings \citep{Higdon96,Amram98,Vorobyov03}, 
radial color gradients \citep{Markum92,Borne97,Korchagin01,Vorobyov01}, 
and radial age distribution of young stellar clusters \citep{Appleton99}.
However, the nature of the pre-collision Cartwheel is still controversial:
it might have been either a gas-rich low surface brightness galaxy 
\citep{Korchagin01} or a regular large galaxy \citep{Vorobyov03}. 
To identify the types of disk galaxies that can serve progenitors to the collisional ring 
galaxies, we should extend the observational and theoretical studies beyond the 
Cartwheel.

The available sample of collisional ring galaxies, 
i.e. galaxies with 
an apparent ring morphology and nearby companions, is relatively small.
However, traces of expanding rings and associated star formation can be
found in otherwise regular galaxies. These features include a shift
between the peaks of radial distributions of OB-associations and HII 
regions in M~51, M~101, and some other nearby 
galaxies \citep{Smirnov78}, recently discovered ring in 
NGC~922 \citep{Wong06}, and ring-like structure in the
inner region of M~31 \citep{Block06}. This evidence suggests that head-on
collisions with satellites are likely to be a more frequent phenomenon in the 
local universe than they were previously thought.

This paper is motivated by the fact that propagating waves of star
formation are expected to leave behind stellar populations, the ages of which should
be characterized by a radial gradient. This gradient should be imprinted 
in the absorption spectra measured at different radial distances from 
the center of a collisional ring galaxy. 
We have chosen Arp~10, a galaxy with an apparent ring morphology,
because of its large size and convenient location for our observations 
(in the Northern Hemisphere). Unlike the Cartwheel, Arp~10 shows clear signs
of the old stellar population in the form of elements of spiral arms and prominent bulge. 
This gives us an opportunity to study the development of ring structures 
in regular spiral galaxies. 
In contrast to previous works \citep{CAM,Marston95,CA96,Bransford98},
we considerably improved the spatial resolution of kinematical 
studies of Arp~10. We also search for the companion that is presumably responsible for the Arp~10 
peculiar ring structure. The new set of observational data allows us to create 
a self-consistent picture of collision and subsequent star formation
in Arp~10.

\section{Observations}
\label{s1}

Observations of Arp~10 were conducted with the 6-meter telescope
at the Special Astrophysical Observatory (Russian Academy of
Sciences) and multi-mode focal reducer SCORPIO
\citep[see][for details]{SCORPIO}.  The SCORPIO allows us to
carry out the broad- and narrow-band photometry, long-slit
spectroscopy, and scanning Fabry-Perot spectroscopy.
The detector is a $2048\times2048$ pixels CCD EEV42-20.

\subsection{Photometry}
\label{s1.1}

\begin{figure}
\plotone{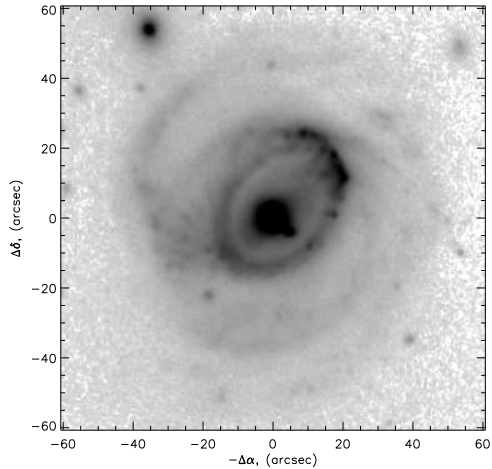}
\caption{B-band image of Arp~10.
\label{f1}
}
\end{figure}

The H$\alpha$ and Johnson-Cousins B- and R-images were
taken on September 21/22, 2003. The pixel size was 0.36 arcsec (with
binning $2\times2$). We obtained several dithered images in each
broadband filter, with a total integration time 600 and 70 sec in
the B and R-bands, respectively.  The mean seeing was 1.6 arcsec.
The images were corrected for bias and flat
field, cosmic ray hits, and sky background. A saturated nucleus in the B- and
R-bands was corrected using short expositions. The B-band image is shown
in Figure~\ref{f1}.

To obtain the H$\alpha$ image corrected for the continuum, 
we used the 15 \AA~ filter centered at 6720~\AA~ for the emission line and 
15 \AA~ filter centered at 6765 \AA~ for the continuum. 
The images in these filters were integrated for 1200 and
600 sec for the H$\alpha$ and continuum, respectively. The
H$\alpha$ image of Arp~10 corrected for the continuum and superimposed onto
the B-band image is shown in Figure~\ref{f2}.


\subsection{Long-slit spectroscopy}
\label{s1.2}

\begin{figure}
\plotone{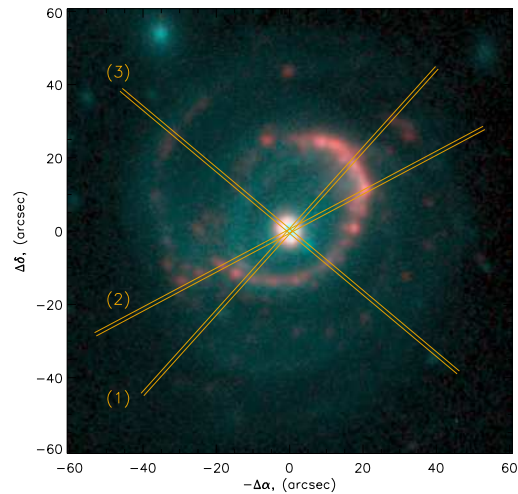}
\caption{$H\alpha$ image of Arp~10 (red) superimposed on the
B-image (blue). Positions of the long-slit cuts are shown by
solid lines. The cuts are labelled according to the text.
\label{f2}
}
\end{figure}

\begin{figure}
\plotone{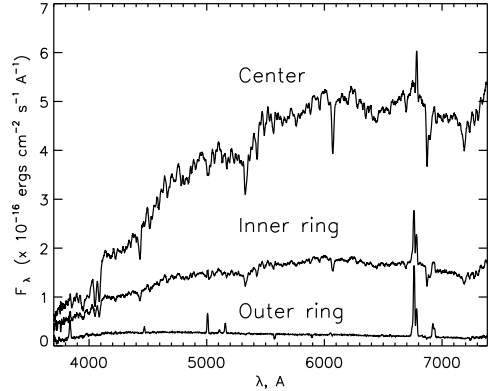}
\caption{One-dimensional spectra of regions within the outer ring, inner ring,
and center of Arp~10 from our long-slit spectroscopy (cut 2).
\label{f_spectra}
}
\end{figure}

We obtained three long-slit spectra of Arp~10 using a 1.0 arcsec wide 
slit. The spectra are taken along three cuts, which pass through the center of the 
galaxy and are shown in Figure~\ref{f2}. 
The summary of spectral observations including the supplementary information 
about the spectral range and resolution, corresponding integration times, and 
position angles of the cuts are given in Table~\ref{t1}. 
The first spectral cut coincides with the major axis of the outer ring 
in Arp~10.  The second cut is drawn at the position angle 118\degr~ and 
covers a larger spectral range with a lower spectral resolution. 
The third cut passes through a small "knot" (presumable satellite-intruder) at
5 arcsec from the center of the galaxy. The spectra were reduced by
the standard procedure (bias subtraction, flat fielding, night sky
subtraction, and wavelength calibration), and then flux-calibrated
with the help of standard stars BD+75~325 and BD+25~4655. 
The dispersion of the spectra along
the cuts is 0.71 arcsec pix$^{-1}$. The seeing was 1.4-1.6 arcsec
during the observations. The one-dimensional spectra of regions within 
the outer ring, inner ring, and center of Arp~10 are shown in
Figure~\ref{f_spectra}.


\begin{table}
\caption{Summary of SCORPIO long-slit observations} 
\label{t1}
\begin{center}
\begin{tabular}{clcccr}
\tableline\tableline
\# &Date & Sp. Range & Sp. Resol. & Exp. time & P.A. \\
   &  & \AA~      & \AA~           & sec       &      \\
\tableline
1 & Sep. 21/22, 2003 & 4150-5890 & 5 &6000 & 138$\degr$ \\
2 & Nov. 02/03     2003 & 3700-7400 & 8 &4800 & 118$\degr$ \\
3 & Aug. 16/17,    2004 & 4040-5600 & 5 &2700 &  50$\degr$ \\
\tableline
\end{tabular}
\end{center}
\end{table}


\subsection{Fabry-Perot Spectroscopy}
\label{s1.3}

The H$\alpha$ velocity field of Arp~10 was obtained via 7320
seconds of integration with the SCORPIO in the scanning
Fabry-Perot Interferometer (FPI) mode with 32 spectral channels. 
Each channel corresponds to 40~km~s$^{-1}$. The formal accuracy of 
our radial velocity measurements obtained from the H$\alpha$ line
fitting by a Gaussian is 10~km~s$^{-1}$. The
typical procedure of reduction of the FPI frames is described in
detail by \citet{Moiseev02}. The spatial sampling of 0.71 arcsec pix$^{-1}$
and velocity resolution of 125 km s$^{-1}$ are achieved in our
frames. The data were smoothed to the spatial resolution
corresponding to a seeing 2.9 arcsec.

\section{Structure of Arp 10 from the B and H$\alpha$ photometry.}
\label{s2}

The broad-band images of Arp~10 indicate the presence of an exponential 
disk and prominent bulge. We fit the R-band image with a model 
exponential disk and bulge \citep[see for details][]{Smirnova06} and
obtain the following parameters: the exponential disk scale length is 11.4
arcsec and exponential bulge scale length is 2.8 arcsec. These scales correspond to 
7.18~kpc and 1.77~kpc, respectively, for the adopted distance of 130 Mpc
($H_0 ~=~ 70$~km~s$^{-1}$~Mpc$^{-1}$). One arcsec corresponds to 0.63~kpc
and the ratio of the bulge-to-disk luminosities is 0.4.

The H$\alpha$ emission of Arp~10 comes mostly from 
two star forming rings, which are clearly seen in Figure~\ref{f2}. However, the H$\alpha$
structure is more complicated than is expected from a simplified
picture, in which a star formation wave propagates from the center toward
the periphery. The outer ring is rather asymmetric, with the H$\alpha$ 
emission coming mostly from its North-Western part (top-right in Figure~\ref{f2}). 
At the same time, the inner ring 
has a concentric shape. A long arc of star-forming regions that traces an outer
spiral arm joins the outer ring in its South-Eastern part (bottom-left
in Figure~\ref{f2}). A similar but a fainter arc is seen on the opposite 
side of Arp~10. 

The structure of the outer disk is best seen in the B-band image (Figure~\ref{f1}). 
Outside the outer ring, at least three spiral arms can be identified, two of them
are revealed in the H$\alpha$ image. The diameter of the outer H$\alpha$ 
ring (28~kpc) is twice smaller than D$_{25}=57$~kpc \citep[according to][]{RC3}. 
This is in agreement with recent multi-band photometry of 15
northern ring galaxies by \citet{Romano07}, who have found that
stellar disks extend outside H$\alpha$ emitting rings in all galaxies of
their sample. Figure~\ref{f1} suggests that the pre-collision Arp~10 might have possessed
a complicated structure typical for spiral galaxies.

\section{Kinematics of Arp 10}
\label{s3}
\begin{figure}
\plotone{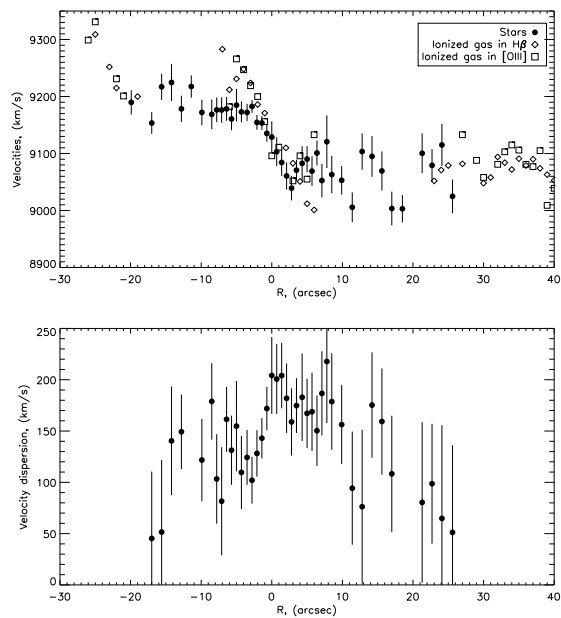}
\caption{Line-of-sight velocities ({\it top} panel) of Arp~10 obtained
with the long-slit
spectroscopy. The ionized gas velocities are shown by diamonds
($H\beta$) and squares ([OIII]). Stellar velocities are designated by
filled circles. Radial distribution of the stellar velocity
dispersion is shown in the {\it bottom} panel.
\label{f3}
}
\end{figure}

In this section we derive the kinematics of Arp~10 using the long-slit
and FPI spectroscopy.
Figure~\ref{f3} shows the H$\alpha$, [OIII], and stellar line-of-sight velocities 
for our long-slit data (top panel) taken along the first cut. The velocities in Figure~\ref{f3} 
are not corrected for inclination. It can be noticed that at some
radii stars rotate faster than the ionized gas, which indicates a
presence of essential non-circular motions in the galaxy. The
bottom panel in Figure~\ref{f3} demonstrates the stellar velocity
dispersion distribution, which is typical for very large
spiral galaxies with high amplitudes of rotation curves \citep{vdKruit86,Bottema93}. 

The two-dimensional H$\alpha$ velocity field of Arp~10 was analyzed
by a tilted-ring technique in the same manner as described by
\citet{Smirnova06}.  A similar technique was applied to the
Cartwheel galaxy by \citep{Amram98} and to Arp~10 by \citet{CA96}.
In contrast to our analysis, Charmandaris \& Appleton used
low-resolution HI data.

\begin{figure}
\epsscale{.7}
\plotone{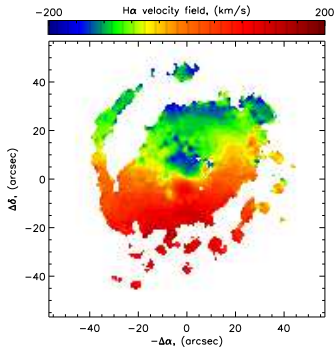}
\caption{Two-dimensional velocity field
in Arp~10 obtained with the scanning FPI.
analysis.
\label{f4}
}
\end{figure}

To analyze the velocity field of Arp~10 shown in Figure~\ref{f4},
we assume that the line-of-sight velocity of the ionized gas
$V_{obs}$ is a combination of the following four components: a
systematic velocity $V_{sys}$, azimuthal velocity in the plane of
the disk $V_{\phi}$, radial expansion velocity in the plane of the
disk $V_r$, and vertical velocity $V_z$:

\begin{eqnarray}
V_{obs} &=&V_{sys}+ V_{\phi} \cos \phi \sin i + 
V_r \, \sin \phi \sin i + \nonumber \\
&+& V_z \cos i.
\label{vel}
\end{eqnarray}
Here $i$ designates the inclination of the galactic plane and $\phi$
denotes the azimuthal angle, which is measured counterclockwise from the
major axis. 
The position angle of
an arbitrary point in the disk p.a. is connected with angle
$\phi$ as $\tan({\rm p.a.} - {\rm P.A.}) = \tan(\phi) \cos(i)$, 
where P.A. is position angle of the major axis (line-of-nodes). 
We split the galactic disk of Arp~10 into 48 tilted circular annuli.
In the case
of a purely circular motion (P.A.=const, $i$=const, $V_r$ = 0,
and $V_z$ = 0), the best fit position angle of the major axis and
inclination are 172$\pm$2\degr~ and 22$\pm$4\degr.  We obtained
these values by masking the inner and outer rings.

\begin{figure}
\epsscale{1.}
\plotone{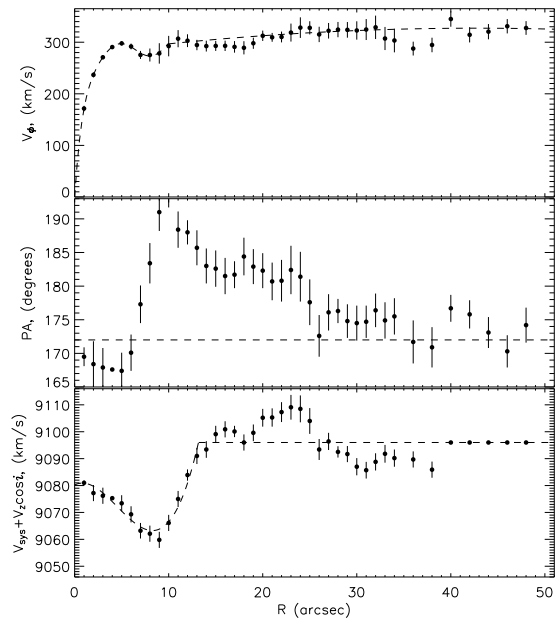}
\caption{Tilted-rings analysis of the velocity field  in Arp~10.
{\it Top} panel: Rotation curve with error bars (circles and vertical 
bars) derived from 2-D velocity field
and model rotation curve (dashed) used to determine residuals 
in Figure~\ref{f4}. 
{\it Middle} panel: position angles at different radii 
and adopted position angle (dashed line).
{\it Bottom} panel: the observed (circles and bars) and model (dashed
curve) sums of systematic and off-plane components of the line-of-sight
velocity in Arp~10.
\label{f5}
}
\end{figure}

\begin{figure}
\epsscale{.6} 
\plotone{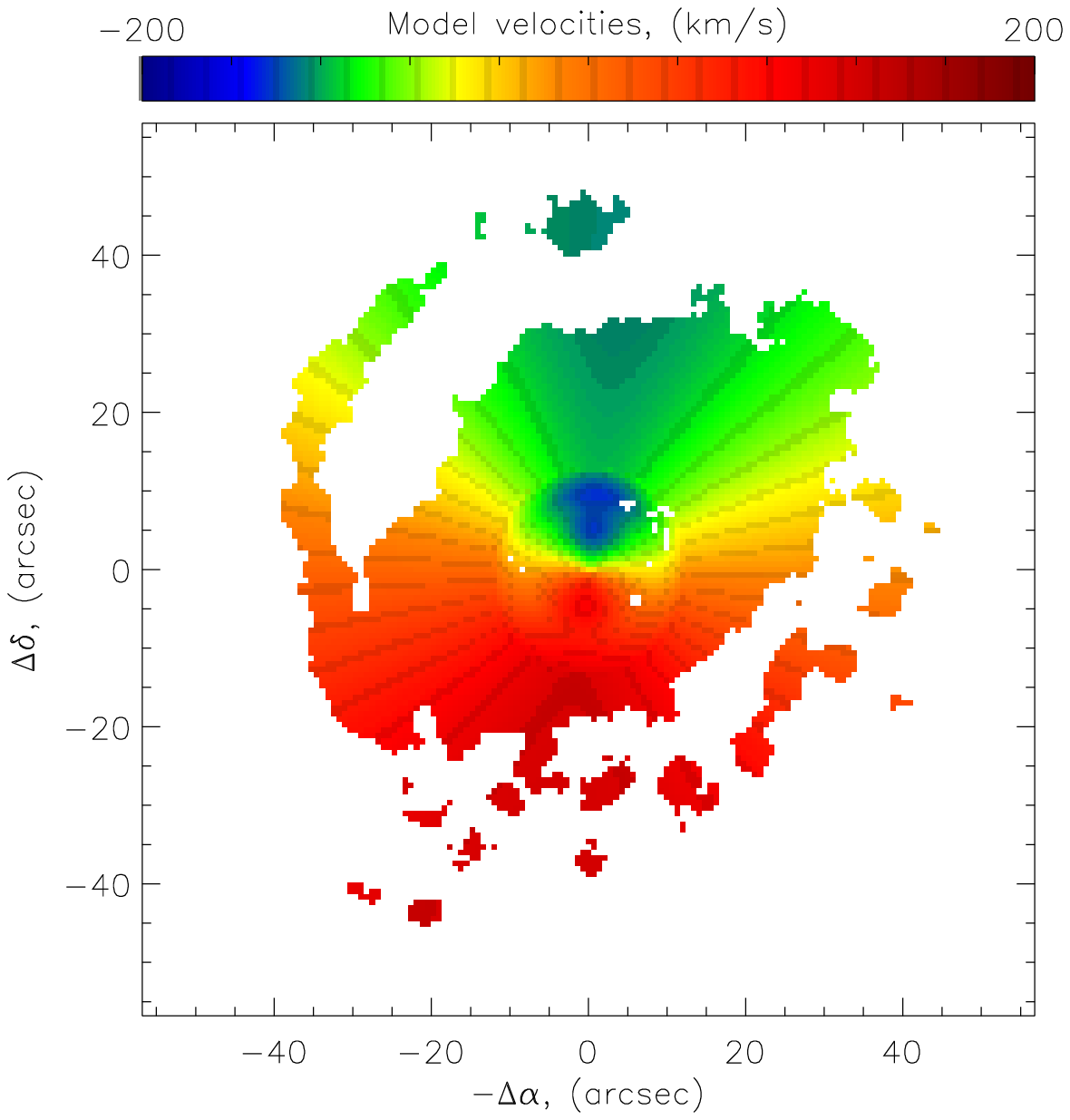}
\plotone{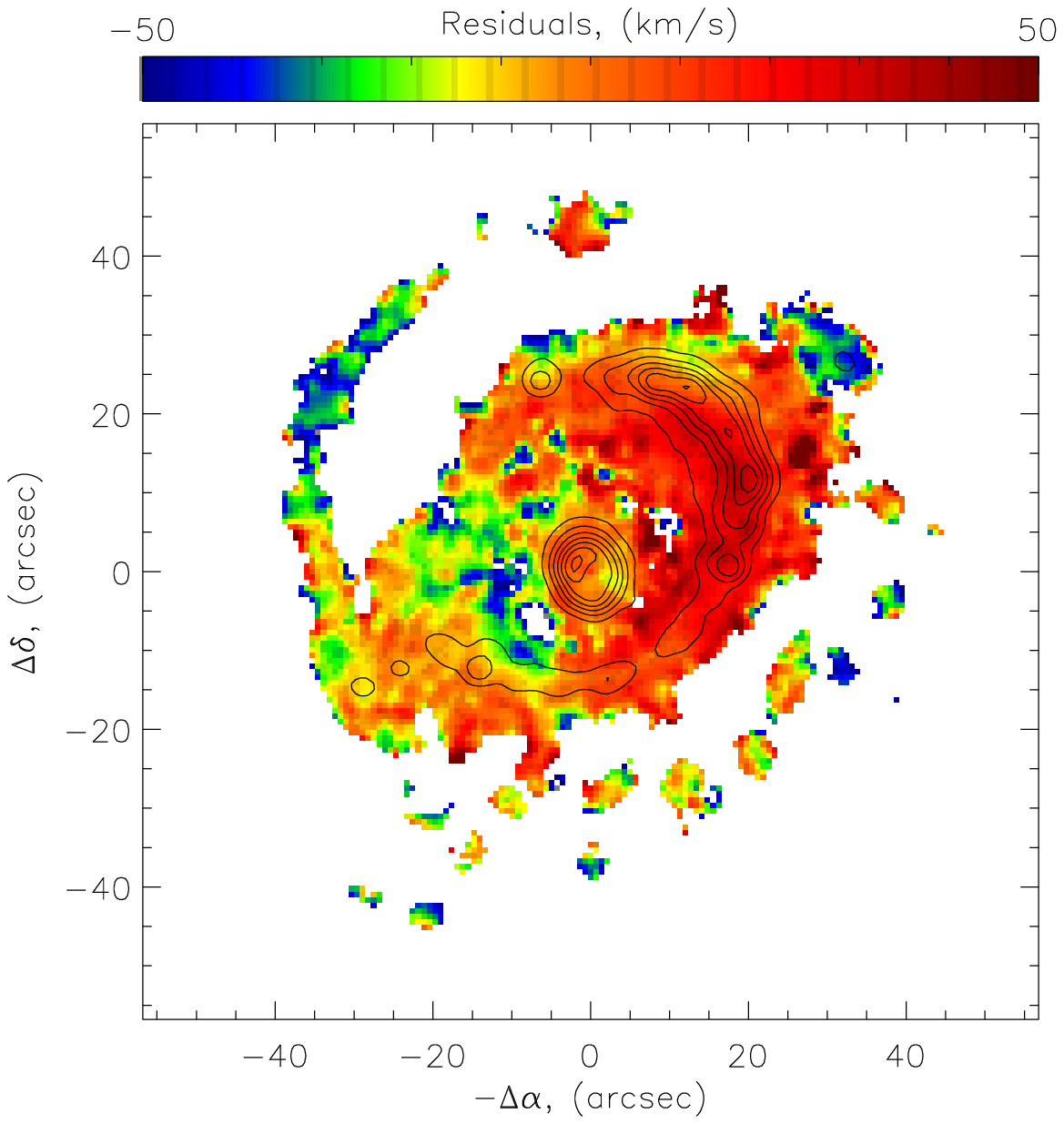}
\caption{{\it Top} panel: the circular velocity field in Arp~10
constructed using equation (\ref{vel}) and values derived from 
the tilted-rings analysis.
{\it Bottom} panel: Line-of-sight velocity residuals in Arp~10.
A contour map of $H\alpha$ emission is 
superimposed by thin black curves.
\label{f4.1}
}
\end{figure}

The mean rotation curve of Arp~10 is derived from our 
two-dimensional H$\alpha$ velocity field and is shown by the
points in the upper panel of Figure~\ref{f5}.  
Next, we use an assumption of quasi-circular
motions and calculate P.A. independently in each elliptical annulus.
The points in the
middle panel of Figure~\ref{f5} show P.A. at different distances
from the center of Arp~10. The points in the bottom panel
represent a sum of the systematic and vertical components of the
line-of-sight velocity. We note that the rotation curve continues
to rise slowly even at large radii (50~arcsec, or 32~kpc). The
dashed lines in all three panels of Figure~\ref{f5} show the values
that are used to construct the model velocity field using
equation (\ref{vel}). This field is presented in the top panel of
Figure~\ref{f4.1}. The model velocity field was subtracted
from the observed velocity field to obtain the residual velocity
field, which is shown in the bottom panel of Figure~\ref{f4.1}.

The kinematic data suggest that the best-fit inclination is
22\degr~ rather than the photometric inclination $49\degr$ used by
\citet{CA96}.  Taking into consideration the asymmetric shape of
the outer ring in Arp~10, this large difference in the values of
inclination is not surprising. \citet{Reshetnikov99} noticed that
Arp~10 would not fit into a general Tully-Fisher relation, if an
inclination of $i$ = 49\degr~ is adopted.  A lower value of
inclination $i$ = 22\degr~ that provides better agreement with the
Tully-Fisher diagram suggests a total mass of Arp~10 to be $7\times
10^{11} M_{\odot}$ within a radius of 30~kpc. The value of the
kinematic position angle of the major axis 172\degr~ is in poor
agreement with the value 125\degr~ derived from the photometric data
\citep{LEDA}. Assuming that the outer spirals in
Figure~\ref{f1} are trailing, we conclude that the galaxy rotates
counterclockwise and its Eastern part (the left half in
Figure~\ref{f2} and Figure~\ref{f4}) is closest to us.

The residual velocity field of Figure~\ref{f4.1} and
radial variations of P.A. and $V_z \cos~i$ in Figure~\ref{f5}
indicate the presence of two main features.

1) Velocities in the inner ring suggest significant vertical
motions of ionized gas with an amplitude of the order of 30
km~s$^{-1}$. The tilted-ring analysis performed for $i$ = 22\degr~
and $V_r$=0 also implies significant vertical velocities $V_z$ in
the vicinity of the inner ring (see middle and bottom panels in
Figure~\ref{f5}).

2) The outer ring expands non-uniformly, with the expansion
velocity in the plane of the disk being larger in the North-Western
part of the ring (top right in Figure~\ref{f4}) than in the
South-Western part. Note that higher expansion velocities in the
outer ring correspond to larger distances from the center. This
implies that different parts of the outer ring has started to
expand at the same time and from the same position in the disk,
which favors the collisional origin of the outer ring in Arp~10.

The radial variations of the kinematic P.A. in the middle panel of
Figure~\ref{f5} is a distinctive feature of radial motions,
according to \citet{Moiseev04} and references therein. Note that we
assume that all non-circular motions in the outer ring occur in the
galactic plane and the azimuthal component of non-circular velocities 
is negligible. Strictly speaking, this is true only for the
regions that lie at the minor axis (i.e. regions with P.A. =
82\degr~ and 262\degr). In all other regions, the vertical and
azimuthal components may contribute to the residuals.

The residual velocities in the inter-ring region may
achieve rather high values. At the same time, the
signal-to-noise ratio (S/N) is quite small there, which makes the interpretation 
of gas motions in the inter-ring region difficult.

\begin{figure}
\plotone{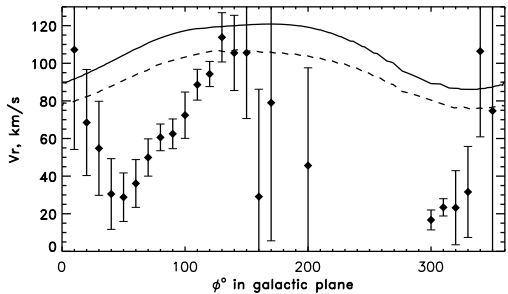}
\caption{In-plane residual velocities (circles and error bars) at the
outer ring of Arp~10 at different in-plane position angles $\phi$ (which
are counted from the south counterclockwise). The solid curve designates the
model expansion velocities obtained in numerical simulations in 
\S~\ref{s7} at 85~Myr after the collsion. The dashed curve is the same but for
105~Myr.
\label{f5.2}
}
\end{figure}

To see how the expansion velocity in the plane of the disk $V_{\rm r}$ varies
along the outer ring, we average $V_r$ over 7 pixel ($\sim$ 5 arcsec) circular regions, which
are regularly spaced along the ring. The resulting distribution of $V_{\rm r}$ along the outer ring 
is shown in Figure~\ref{f5.2} as a function of
angle $\phi$. Here $\phi$ is measured counterclockwise in the plane of Arp~10
from its south direction. The fastest expanding part 
($\approx 110$~km~s$^{-1}$) of the outer ring  at $\phi \sim$ 130\degr -- 150\degr~ 
corresponds to the highest surface brightness
in H$\alpha$. The opposite side of the ring at $\phi \sim$ 310 - 330\degr~ expands 
considerably slower (up to 30 km s$^{-1}$). 

\section{Long-slit spectra of Arp 10}
\label{s4}

Although we obtained three long-slit radial cuts,
only the second cut (hereafter, cut~2) is well suited for the analysis of emission and
absorption lines because of its wide spectral coverage and best S/N ratio.

\subsection{Emission lines}
\label{s4.1}

The most intensive emission lines are seen in the
star-forming rings of Arp~10. Cut~2 in Figure~\ref{f2} (P.A.=118\degr) covers the regions of
most intensive emission and ranges from 3700 \AA~ to 7400 
\AA, which allows us to estimate extinction 
and heavy element abundances along the cut. 
To estimate the extinction, we use the H$\alpha$/H$\beta$ line ratio.
For the correct extinction and abundance analysis, we have to take the
absorption lines into account. 

In order to obtain the emission line fluxes, we fit the
observed spectrum around several selected spectral lines 
([OII]$\lambda \lambda$3727,3729, $H\beta$, [OIII]$\lambda$4959, and
[OIII]$\lambda$5007) with  a combination of model continuum, absorption lines, 
and one Gaussian emission line. In the case of blended lines 
(H$\alpha$, [NII]$\lambda$6548, [NII]$\lambda$6584), three Gaussian emission lines 
were used. The free parameters were
the width and height of emission lines, and the height of continuum. The
best-fit parameters were used to obtain the emission line
fluxes. These fluxes at different distances from the center of Arp~10
are given in Table~\ref{t2}.

The extinction in the V-band ($A_V$) from HII regions located at different radii is estimated 
using the parameterization of \citet{Charlot00}. 
The results are shown in Table~\ref{t3}. The extinction can be determined 
only in star-forming rings: it attains a mean value of 0.32~mag in the inner ring
and increases toward the outer ring, reaching a maximum value of 4.0~mag 
in the external part of the outer ring. This is in general agreement with a propagating
star formation scenario -- massive young
stars destroy dust and decrease extinction in the wake of a propagating wave of star formation.
The observed emission fluxes in Table~\ref{t2} were corrected for extinction
and then were used to evaluate the oxygen abundance in Arp~10. 
The temperature-sensitive line [OIII]$\lambda$4363 is not seen in our spectra.
Instead, we follow an approach described by \citet{PT05},
which uses the excitation parameter $P$ based on the
[OII]$\lambda\lambda$3727,3729 to [OIII]$\lambda\lambda$4959,5007 relation 
$R_{23}$:

\begin{equation}
12 + \log(O/H) ~=~ \frac{R_{23} + 726.1 + 842.2 P + 337.5 P^2 }
{85.96 + 82.76 P + 43.98 P^2 + 1.793 R_{23}} ~~~,
\end{equation}

\noindent
where $P = R_3 / (R_2 + R_3)$, $R_2 = I_{[OII]\lambda\lambda3727,3729} 
/ I_{H\beta}$, $R_3 = I_{[OIII]\lambda\lambda4959,5007} / I_{H\beta}$,
and $R_{23} = R_2 + R_3$.

The derived abundance of oxygen will be applied in the spectral index
modeling in \S~\ref{s5}.  The shape of absorption spectral lines
depends on abundances of heavy elements.  In turn, the emission
line fluxes (and hence the extinction and abundances) depend on the
subtracted absorption spectrum. To determine abundances and
extinction, we use the following procedure. We estimate
extinction/abundances from emission lines and use the resulting
values to derive the model absorption spectrum. The resulting
spectrum in turn is used to derive a new approximation for the
emission line fluxes and, by implication, for
extinction/abundances.  This procedure is iterated until
convergence.

The resulting values of $P$ and oxygen abundances 12 + log(O/H) are
given in Table~\ref{t3}. These values agree with a mean value of 12
+ log(O/H) = 8.6 estimated by \citet{Bransford98} for two HII
regions in the outer ring of Arp~10. However, when a method of
evaluating oxygen abundances from \citet{Bransford98} is applied to
our emission line fluxes, the obtained
oxygen abundances exceed the solar value at all radii. This is
unlikely for Arp~10 due to its presumably low surface density
nature (see \S~\ref{s7.1}). We note that the radial oxygen
abundance gradient in Arp~10 is very small, with a mean oxygen
abundance of 8.55 and 8.64 in the inner and outer rings,
respectively. This implies a negligible radial metallicity gradient
in young stars as well.


One of the products of our analysis is the radial distribution of
extinction-corrected fluxes in H$\alpha$ and H$\beta$, which are
shown in Table~\ref{t3}.  The H$\alpha$ fluxes were corrected for
[NII]-emission by deblending. 
We note that the [NII]/H$\alpha$ flux ratio is remarkably constant along the radius of Arp~10. 
Table~\ref{t3} indicates that [NII]$\lambda$6584/H$\alpha$ = 0.37 $\pm$ 0.01 and
([NII]$\lambda$6548 + [NII]$\lambda$6584)/H$\alpha$ = 0.48 $\pm$ 0.05. 
These values are typical for quiescent regions in regular galaxies 
rather than for regions where strong shock waves are seen \citep{KennicuttKent83,Moiseev00}. 
It suggests that propagating gas density waves in Arp~10 do not 
generate strong shock waves.


\subsection{Spectral indices}
\label{s4.2}

The equivalent width of absorption lines in galaxies depends on
a contribution of stellar populations of different ages.
In case of Arp~10, we assume that
stellar populations consist of old background stars and young
stars. The latter were born after a passage of a star formation wave associated
with an expanding gas density wave.
We apply several population synthesis codes to construct model spectra and
estimate the relative amount of old and young stars at a specific radius
in Arp~10. The propagating wave of star formation in Arp~10 is approximated 
by a sequence of single-burst events. Spectra of single-burst populations are functions of
age, metallicity, and initial mass function (IMF). 
We assume that the IMF has a universal nature  and does not depend on a particular
location in Arp~10.

In order to compare model and observed spectra, the latter should be
corrected for internal extinction. The values of V-band extinction from HII
regions estimated in \S~\ref{s4.1} may differ considerably from the
interstellar extinction in other regions of Arp~10. Fortunately, most
spectral indices do not depend on extinction, since they are defined in
relatively narrow bands of the spectrum.  We use index definitions by
\citet{Worthey94}, \citet{Worthey97}, \citet{Trager98}, and
\citet{Gorgas99}. Out of 31 indices, 27 can be measured in our spectral
range.

To find the indices that can effectively serve as indicators of age and metallicity,
we use three population synthesis codes: STARBURST99  \citep[the latest version is
based on][]{Vazquez05}, GALAXEV \citep{BC2003},
and Worthey's code \citep{Worthey94}. The STARBURST99 code produces
high-resolution synthetic spectra with dispersion 0.3 \AA~pixel$^{-1}$ in
our spectral range. It allows us to smooth the model spectra down to our
resolution of 5 \AA~ and 8 \AA. The STARBURST99 code allows to obtain
spectra of a single-burst population and to choose parameters for 
a given IMF. We use the Kroupa IMF \citep{Kroupa01} and
Padova set of stellar models. The spectral indices were calculated from 
the obtained high resolution synthetic spectra.
The GALAXEV code provides spectral indices for a set of
ages, metallicities, spectral resolutions, 
and stellar velocity dispersions. We choose the Chabrier IMF \citep{Chabrier03}, 
which is similar to the Kroupa IMF, and Padova evolutionary tracks. 
The third population synthesis code \citep{Worthey94} is considered
here only for illustrative purposes because of its incompleteness
at ages of the order of 100 Myr.

\begin{figure}
\plotone{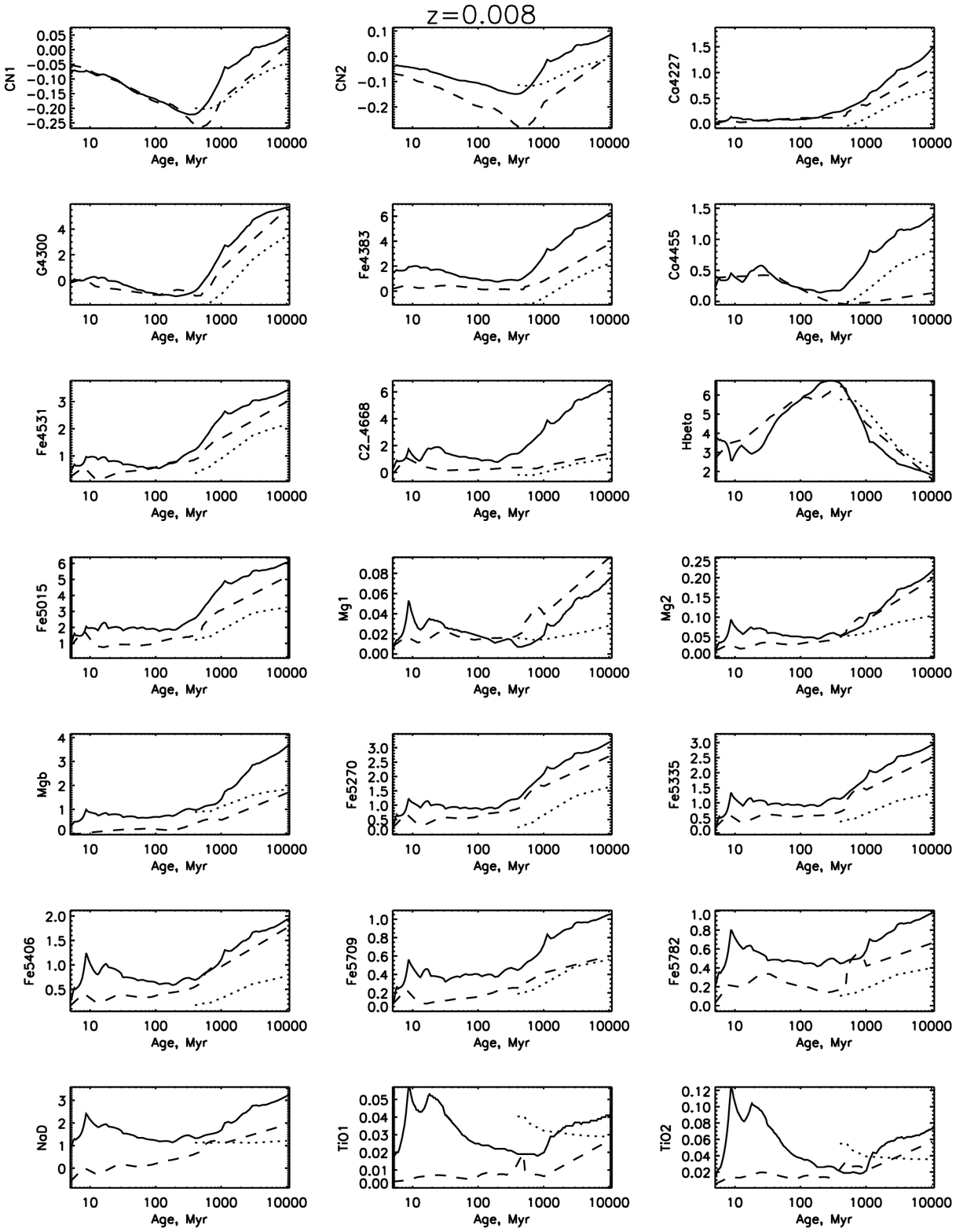}
\caption{Eighteen spectral indices derived for a single-burst stellar 
population with metallicity  $z=0.008$  (assuming that the solar metallicity is 0.02)
at different ages in three models: 
solid curves -- GALAXEV \citep{BC2003},
dashed curves -- STARBURST99 \citep{SB99},
dotted curves -- Worthey \citep{Worthey94}.
\label{f6}
}
\end{figure}

\begin{figure}
\plotone{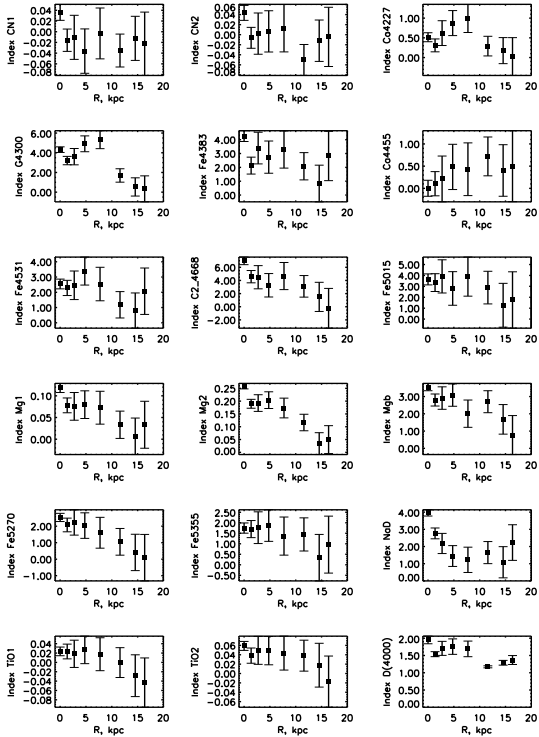}
\caption{Eighteen spectral indices (the same as in Figure~\ref{f6})
observed in Arp~10 at different distances from the center for
long-slit cut 2. 
\label{f6.2}
}
\end{figure}

Figure~\ref{f6} shows how the spectral indices obtained with 
STARBURST99 (dashed lines), GALAXEV (solid lines), and Worthey's code (dotted lines)
change with age of a single-burst population. The metallicity is fixed in all models and
equal to 0.008 (assuming 0.02 for the solar metallicity), which is close to a mean value of 
metallicity of young population derived for Arp~10 in
\S~\ref{s4.1}. We note that the indices derived with the Worthey's code 
do not extend below 400~Myr. STARBURST99 and GALAXEV show general agreement,
although a systematic offset in the values of indices is seen between the two codes.
For our analysis, we need spectral indices that show a significant monotonous variation 
with age. As is seen in Figure~\ref{f6}, not all indices change monotonously 
between 0~Myr and 500~Myr, which is an expected range of ages of young stars in 
collisional ring galaxies. Some indices vary negligibly with age.
An additional uncertainty is introduced by a dependence of spectral indices on metallicity. 
In principle, the hydrogen line indices based on the strength of H$\beta$, $H\gamma$, 
and $H\delta$ lines are the best age indicators. Unfortunately, a presence of emission 
lines makes it impossible to utilize them.
Another age-sensitive index $D4000$ has to be used with caution, 
because it is defined on a 500 \AA~ bandpass in the blue part
of spectrum and thus depends on the poorly known value of internal extinction. 
For instance, an extinction of $A_V = $ 1 mag will decrease $D4000$ by 7\% \citet{Gorgas99}. 
Three iron-sensitive indices (Fe5406, Fe5709, and Fe5782)
are not good for the analysis, because their bandpasses interfere with 
strong night sky lines in our spectra.
Finally, we accept 18 spectral indices for the
further analysis of Arp~10. The observed radial distribution of these
indices is shown in Figure~\ref{f6.2}

A brief inspection of Figure~\ref{f6.2} and comparison of Figure~\ref{f6.2} with 
Figure~\ref{f6} show that the indices are sensitive to the presence of
young stars in the inner and outer rings. Systematically smaller values 
of Fe, Mg, Ca, Ti, and D4000 indices are seen in the both rings. 
Since Arp~10 is unlikely to have a similar systematic variation in metallicity,
the observed variation of indices should be interpreted as a presence of young stellar population.

\section{Spectral index modeling of propagating star formation in Arp~10}
\label{s5}
In this section we make predictions about metallicities
and ring propagation velocities in Arp~10 by fitting the radial distribution of
model spectral indices to that of the observed ones. 

\subsection{Model description}
\label{s5.1}
We assume for simplicity that the stellar disk of Arp~10 consists of young
and old stellar populations. 
The old stellar population has an age of either
10~Gyr or 5~Gyr. The former value is typical for elliptical galaxies without ongoing star formation
and early-type spiral galaxies. The latter value
is typical for blue low surface brightness and for late-type
galaxies. These values are chosen to represent two limiting ages of old stellar populations in galaxies. 
The old stellar population in our model consists of an exponential disk and bulge, 
the radial scale lengths of which  ($h_{old}=7.2$~kpc and $h_b=1.77$~kpc, 
respectively) were obtained from photometry in \S~\ref{s2}.  
The radial surface density distribution of the old stellar population is described as 
\begin{equation}
\Sigma^{old} = \Sigma_{0}^{old} (17.67 \exp(-r/h_b) + \exp(-r/h_{old})),
\end{equation}
where $\Sigma_{0}^{old}$ is derived in \S~\ref{s7} to be 300~$M_\odot$~pc$^{-2}$ and
the bulge-to-disk ratio of their central surface densities is 17.67.

The young population emerges after a passage of two circular gas density waves, which
propagate from the center of Arp~10 toward its periphery and 
trigger star formation. We assume constant propagation velocities of the outer and
inner waves (representing the outer and inner rings) and denote their values
by $V_{outer}$ and $V_{inner}$, respectively.
We assume that the radial distribution of metallicity of both young and old stars 
can be approximated by a linear function of radius. In the following text, 
we normalize metallicities (derived from the spectral index modeling) to the 
solar value ($z=0.02$) and denote the resulting relative metallicities as $Z$.
Using the oxygen abundance estimates in \S~\ref{s4.1} and assuming that the 
fractional abundances in Arp~10 are similar to those at the Sun, we assume
the relative metallicity of young stars in the center of Arp~10 to 
be $Z_{0,y}=-0.3$~dex. The metallicity gradient of young 
stars $(dZ/dr)_{y}$  is negligible and we set its value to zero.

We find the metallicity of old stars $Z_{0,old}$ in the center of Arp~10 and metallicity 
gradient $(dZ/dr)_{old}$ using  the least-square fit of model spectral 
indices to the observed indices in the inner 0.7~kpc and in the outer region between 16.3~kpc and 18.8~kpc. 
This can be done because the region outside the outer ring consists of the old stellar population only. 
The center of Arp~10 does contain a certain fraction of the young stellar 
population, but its contribution is expected to be small due to the presence of a 
high surface brightness bulge. First, we use all the indices shown 
in Figure~\ref{f6.2}, except for 
CN1, CN2, and NaD which were found to worsen the fit. 
This yields $Z_{0,old}$ = $-0.03 \pm 0.11$~dex and $(dZ/dr)_{old}$ = -0.10 $\pm$ 0.06 dex~kpc$^{-1}$. 
If we consider only Ca, Mg, and Ti-dominated indices, we 
obtain $Z_{0,old}$ = 0.08 $\pm$ 0.09~dex and $(dZ/dr)_{old} = -0.08 \pm 0.12$~dex~kpc$^{-1}$.
When the iron-dominated indices (G4300, Fe4383, Fe4531, and Fe5270)
are considered, we found that $Z_{0,old} = -0.35 \pm 0.06$~dex and 
$(dZ/dr)_{old} = -0.08 \pm 0.02$ dex~kpc$^{-1}$.
The values of $(dZ/dr)_{old}$ derived using different sets of indices are similar to each other, 
whereas the values of $Z_{0,old}$ obtained from the 
iron- and $\alpha$-element dominated indices are 
significantly different. The values of the abundance gradients
are typical for spiral galaxies 
\citep[from -0.18 to 0.0, according to ][]{Zaritsky94,vanzee98,Rolleston00}.

We assume that the radial surface density distribution of young stars can be approximated 
by an exponential function $\Sigma^{y} = \Sigma_0^{y} \exp(-r/h_{y})$. 
The central surface density of young stars $\Sigma_0^{y}$ is expressed in term of the 
central surface density of gas  $\Sigma_0^{y} = \epsilon \Sigma_{\rm g0}$,
where $\Sigma_{\rm g0}=22~M_\odot$~pc$^{-2}$. 
The coefficient $\epsilon$ is a free parameter, which
characterizes the gas to stars conversion efficiency and possible variation of the gas
surface density.
We denote $h_{y}$ to be the radial scale length of young stars.
The value of $h_{y}$ is set to be equal or larger than the
exponential scale length of the old stellar disk (7.2 kpc), since
the gas disk usually has a shallower radial surface density profile
than the stellar disk, particularly, in Arp~10 \citep{CA96}.

We calculate the model spectral indices $I_{\rm m}$ as 
\begin{equation}
I_{\rm m}(r) = \frac{A_{\rm old} + A_{\rm y}}{C} ,
\end{equation}
if the index is defined in units of equivalent width \citep{Worthey94} or as
\begin{equation}
I_{\rm m}(r) = -2.5 \log \frac{B_{\rm old} + B_{\rm y}}{C},
\end{equation}
if the index is measured in magnitudes. The coefficients 
$C$, $A_{\rm old,y}$, and $B_{\rm old,y}$ are determined as
\begin{eqnarray}
A_{\rm old,y} &=& \Sigma^{\rm old,y}(r) 
\L_V(T_{\rm old,y},Z_{\rm old,y}) \times \nonumber \\
&\times& I_{\rm old,y}(T_{\rm old,y},Z_{\rm old,y}), \nonumber \\
B_{\rm old,y} &=& \Sigma^{\rm old,y}(r)
\L_V(T_{\rm old,y},Z_{\rm old,y}) \times \nonumber \\
&\times& 10^{-0.4 I_{\rm old,y}(T_{\rm old,y},Z_{\rm old,y})}, \nonumber \\ 
C &=& \Sigma^{\rm old}(r) \L_V(T_{\rm old},Z_{\rm old}) + \nonumber \\
&+& \Sigma^{\rm y}(r) \L_V(T_{\rm y},Z_{\rm y}). 
\label{eqspec1}
\end{eqnarray}
The superscripts/subscripts  ``old'' and ``y'' refer to old and young stellar populations, 
respectively. The V-band luminosity of a single-burst population
with relative metallicity $Z_{\rm old}$ or $Z_{\rm y}$ and age
$T_{\rm old}$ or $T_{\rm y}$ is denoted as $L_{\rm V}$.
Spectral indices of a single-burst stllar population are denoted as $I$.

Our model has four free parameters: the gas-to-star conversion 
efficiency $\epsilon$, scale length $h_{y}$, $V_{outer}$, and $V_{inner}$.
Chi-square values $\chi^2$ are calculated for all selected indices
using their model and observed values (along with their
uncertainties) at different radii. An integrated value of $\chi^2$ indices is found
by summing the individual $\chi^2$ indices. We determine the free
parameters of our model by minimizing the integrated $\chi^2$
value. We used the Levenberg-Marquardt optimization algorithm.
The wave propagation velocities may be non-isotropic.
Therefore, a best match between between model and observed spectral indices was obtained
separately for the lower-left
and upper-right halves of cut~2 in Figure~\ref{f2}.

\subsection{Results of spectral modeling}
\label{s5.2}

A comparison of Figures~\ref{f6} and \ref{f6.2} indicates that
synthetic spectral indices obtained with STARBURST99 often fall
outside the range covered by the observed indices.  This effect
does not depend on the S/N ratio of the spectra or uncertainty in
the indices. One of the most prominent examples of this discrepancy
is seen for Mg~b: no agreement between the model and observations
can be achieved for any parameters of stellar populations.
At the same time, the model and observed radial distributions of the
Mg~b index are nearly parallel to each other.  We find that
synthetic indices obtained with GALAXEV reproduce the range
covered by observed indices better than STARBURST99. 
Therefore, we use the GALAXEV code to obtain synthetic indices in this paper.

\begin{figure}    
\plotone{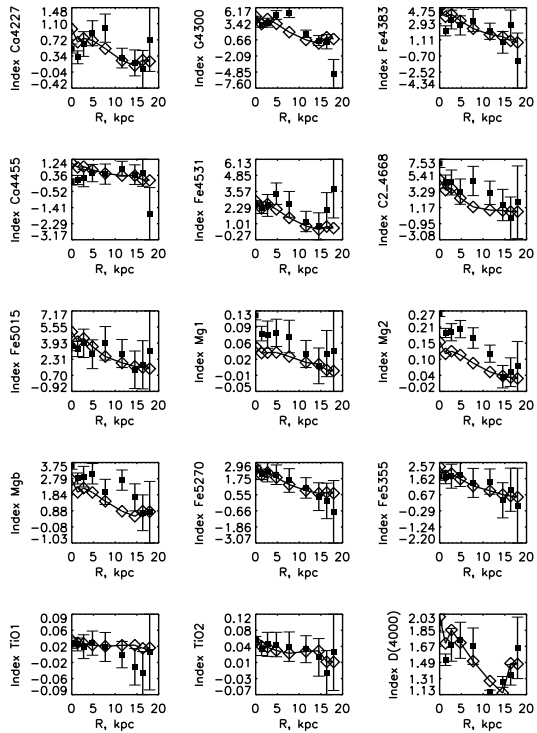} 
\caption{Fifteen model (solid curve and open diamonds)
and observed (filled squares and error-bars) 
spectral indices 
at different distances from the center of Arp~10.
\label{f7} 
}
\end{figure}

First, we consider fifteen observed spectral indices shown in Figure~\ref{f6.2}.
The best match between these indices and model indices is achieved for
the following values of free parameters: $V_{outer}$ =
180$^{+215}_{-55}$~km~s$^{-1}$, $V_{inner}$ = 
46$^{+45}_{-15}$~km~s$^{-1}$, $h_{y}$ = 14 kpc. 
Although the uncertainty in the propagation velocity of the outer ring
is rather large, the value of $V_{outer}$ suggest a very fast expansion
of the outer ring.
The radial distributions of the model (open
diamonds) and observed (filled squares with error bars) indices
obtained along the upper half of cut~2 are shown
in Figure~\ref{f7}.

Second, we consider the spectral index D4000 alone.
Although the D4000 index may be affected by an poorly known value of internal extinction,
mean gas surface density (see \S~\ref{s7}) implies a
rather small mean value of $A_{\rm V}=0.2$~mag \citep{Bohlin78}.
This value of extinction may decrease the D4000 index
by only 1.5\%. The best match between the observed and model 
D4000 indices is obtained for the following values:  $V_{outer}$ =
180$^{+365}_{-50}$~km~s$^{-1}$, $V_{inner}$ =
58$^{+16}_{-12}$~km~s$^{-1}$, and $h_{\rm y}=65$~kpc. 
Finally, we consider two more combinations of spectral indices:
Ca, Mg, Ti, D4000 and Fe, D4000. 
The resulting free parameters also suggest a large expansion velocity for the outer ring 
and large value of $h_{y}$. 

The free parameter $\epsilon$ takes a best-fit value of approximately 0.4. 
This value is a factor of four larger than a typical efficiency of gas-to-stars conversion in 
star-forming regions. A larger value of $\epsilon$ suggests that either the surface density of gas is 
larger than it was assumed and/or the efficiency of star formation
is higher. A higher star formation efficiency was indeed reported for 
the Cartwheel ring galaxy by \citet{Vorob03}. 
The above spectral modeling assumed that the old stellar population had a mean age of 
5~Gyr. A larger value of 10~Gyr yields only a slightly
worse agreement between the model and observed indices.

\section{The intruder galaxy}
\label{s6}

One of the candidates for the intruder galaxy was investigated by \citet{CA96}. They found
that a companion located 60 arcsec North-East of the main galaxy is a distant 
background galaxy. They also suspected another candidate for the intruder galaxy
which can be seen as a small "knot" 5 arcsec to the South-West of the nucleus of Arp 10. 
We consider this possibility and make our third long-slit cut through this
object (hereafter, the companion). The companion has a typical galactic spectrum 
with absorption Balmer lines.

\begin{figure}
\plotone{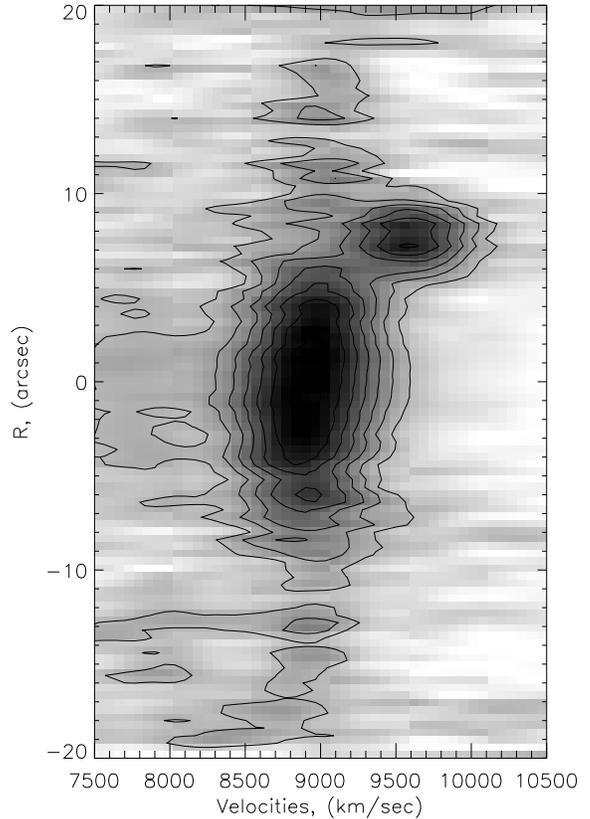}
\caption{Position-Velocity diagram for stellar line-of-sight
velocity distribution (grey-scale) and contour map for the central 20 arcsec in Arp~10.
A difference in the radial velocities of Arp~10 and companion is approximately 
480~km~s$^{-1}$.
\label{f10}
}
\end{figure}

The radial velocity of the companion exceeds 
the systematic velocity of Arp~10 by 480 km s$^{-1}$.
The position-velocity diagram made with the help of absorption lines 
in a wavelength band from 5000~\AA~ to 5400~\AA~
is shown in Figure~\ref{f10}. The line-of-sight velocity distribution
was obtained via a cross-correlation of the object spectra against
a twilight-sky template observed at the same night.
It is clearly seen that the companion has a noticeable velocity 
dispersion $240$~km~s$^{-1}$, which is comparable to that of Arp~10.
We consider isodenses that outline both Arp~10 and companion in Figure~\ref{f10}
and estimate a maximum range of distances $\Delta r$ and velocities $\Delta V$ 
covered by these isodenses for Arp~10 and companion. 
The kinematic masses of Arp~10 and companion are then estimated using 
the following formula $M = \Delta V^2 \, \Delta r / G$. The resulting 
target-to-companion mass ratio is about 4, which corresponds to the total
mass of companion about $2\times 10^{11}~M_\odot$.
Of course, only the inner part of Arp~10 was included in the analysis.
Arp~10 appears to be much more extended than the companion, yet the faint parts of the latter might 
be difficult to observe.

We note that the target-to-companion mass ratio can be larger if the companion is a
non-virialized stellar system. On the other hand, the companion may lose 
a substantial part of its mass during the collision \citep[about
1/3 in a model of][]{Lynds76}, hence the pre-collision mass of the 
companion can be larger. 

Our B-, R-band and H$\alpha$ images of Arp~10 do not resolve the structure of 
the companion's outer regions. Nevertheless, indirect evidence of 
the companion type comes from the position-velocity
diagram in Figure~\ref{f10}, our H$\alpha$ image, and HI map of
\citet{CA96}. Neither significant amount of HI
nor ionized gas are seen in the region of the "knot". 
At the same time, the stellar velocity
dispersion is quite large and the companion appears to be
compact. A visual inspection of Figure~\ref{f10} indicates that the companion has a size of 
5 arcsec or 3~kpc. This suggests
that the companion is located behind the galactic plane of Arp~10
and is shielded by the disk of the main galaxy.

We consider several possibilities for the type of the companion, which might have been 
a disk, elliptic, or irregular galaxy before the collision. 
Disk galaxies are expected to contain a modest amount
gas, which might have been stripped off after the collision. 
Indeed, a gas condensation can be seen in
the HI map of Arp~10 published by \citet{CA96} (see their figure 4, top left panel
and figure 5). A relatively high line-of-sight velocity of this gas clump suggests that it may
not belong to Arp~10. The location of the clump does not coincide 
exactly with the companion's nucleus, as is indeed expected if the companion is not traveling along the line of sight. 

A large total mass of the companion and regular shapes of its isophotes 
argue against a supposition that
the companion was an irregular galaxy before the collision.
The characteristic age of about 5~Gyr for the companion's stellar population (see \S~\ref{s6.1})
implies that the companion was a spiral rather than an elliptical galaxy.

\subsection{Constraints on the companion's age from its spectrum}
\label{s6.1}

\begin{figure}
\plotone{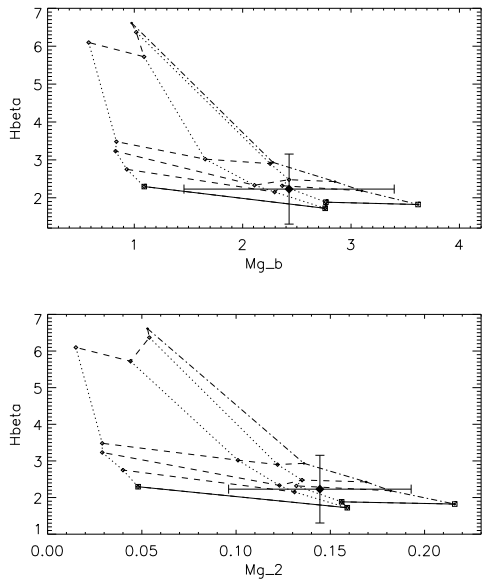}
\caption{$H\beta$ versus Mg~2 ({\it top}) and
$H\beta$ versus Mg~b ({\it bottom}) diagrams. Observed spectral indices
of the companion are designated by filled diamond with error bars.
The dotted lines connect spectral indices of a single-burst 
populations with equal
metallicities of 0.008, 0.001, and 0.0008. The indices for the solar
metallicity $z=0.02$ are connected by the dash-dotted line. 
Spectral indices of the single-burst populations with equal 
ages of 0.4~Gyr, 2~Gyr, 3~Gyr, and
5~Gyr are connected by the dashed lines. The solid line with squares
connects the indices for the age of 10~Gyr.
\label{f11}
}
\end{figure}

The spectrum of the companion shows no emission lines. This allows us
to use the H$\beta$, H$\gamma$, and H$\delta$ spectral indices to
estimate the mean age of stellar population in the companion.  In
Figure~\ref{f11} we show H$\beta$ versus Mg~b (upper panel) and
H$\beta$ versus Mg~2 (lower panel) relations obtained from
single-burst models with different ages and metallicities.  The model
indices were obtained using the GALAXEV package \citep{BC2003} and
adopted to our spectral resolution and velocity dispersion. The
dotted lines in Figure~\ref{f11} connect the indices with equal
metallicities of 0.008, 0.001, and 0.0008.  The indices with a solar
metallicity $z=0.02$ are connected by the dash-dotted line. On the
other hand, the indices with equal ages of 0.4~Gyr, 2~Gyr, 3~Gyr, and
5~Gyr are connected by the dashed lines. The solid line with squares
connects the indices with an age of 10~Gyr.

The observed indices of the companion are shown by the filled
diamonds with the error bars in each panel of Figure~\ref{f11}. 
As can be seen, a single-burst model with an age of 5~Gyr and metallicity $z=0.008$    
yields model indices that are similar to those of the companion.
An estimated mean age of 5~Gyr for the companion's stellar population implies
that the companion was a spiral rather than an elliptical galaxy.

Now, we model the $H\beta$ spectral index under an assumption of continuous
star formation. More specifically, we assume that the star formation rate
declines exponentially with time as $SFR \sim \exp(- \beta \cdot t)$, where
$1 / \beta$ is the characteristic time scale for star formation.  In
elliptical galaxies $1 / \beta$ is less than 1~Gyr \citep{Sandage86},
whereas $1 / \beta \rightarrow \infty$ in irregular systems \citep{GHT84}. 
The resulting H$\beta$ spectral index is
\begin{equation}
I_{H\beta} = \int^T_0 I(t,z) L_V(t,z)\, SFR\, dt,
\end{equation}
where $I(t,z)$ is the value of the $H\beta$ index for a single-burst stellar population
with age $t$ and metallicity $z$, V-band luminosity $L_V$ is the same as in equation
(\ref{eqspec1}), and $T$=15 Gyr.
For the estimated metallicity of the companion $z=0.008$, the observed value
H$\beta=2.25$ \AA~ corresponds to $1 / \beta \approx 4$~Gyrs, which is
typical for Sab galaxies \citep{McGaugh97}.


The foreground young population of Arp~10 and possible
(yet undetected) Balmer line emission from the companion may
change the values of Balmer indices. This may alter the
obtained values of mean age and characteristic time scale for star formation in the companion.
However, this effect is expected to be small due to a lack
of gas in the companion. Moreover, the
emission-independent indices Mg~b and Mg~2 in Figure~\ref{f11} suggest a similar mean age 
of the companion.

\subsection{Synchronized motion of the outer ring and companion}
\label{s6.2}

The companion has a very high velocity of 480~km~s$^{-1}$ relative
to Arp~10.  A mean rotation velocity of Arp~10 is about
300~km~s$^{-1}$ (see Fig.~\ref{f5}), which implies that the escape
velocity for Arp~10 is about 420~km~s$^{-1}$ and the companion may
not be gravitationally bound. However, a near-central collision
suggests that Arp~10 may have a massive dark matter halo that
extends far beyond 50 arcsec. In this case, the companion would be
marginally bound to Arp~10. This might also allow for a much
earlier encounter that might have generated the faint ripples noted
by \citet{CA96}. The numerical modeling in \S~\ref{s7} indicates
that approximately 85~Myr has passed since the collision and the
current distance to the companion from the center of Arp~10 is
about 50 kpc. At the same time, the projected distance from the
companion to the center of inner ring of Arp~10 is about 3 kpc.
Hence, the direction of the companion motion after the collision
nearly coincides with the line of sight.

The numerical modeling in \S~\ref{s7} suggests that 
the companion might have changed significantly its direction of motion after
the encounter. 
Position angles of the companion and the brightest part of the outer ring are different.
The rotation curve of Arp~10 indicates that it takes approximately 170~Myr for the brightest part 
of the outer ring to complete a circle. This implies
that the ring has made about half a turn since its emergence 85~Myr ago.
We reconstruct a synchronized motion of the companion and outer ring as follows. 
The companion passed through the disk of Arp~10 about 3~kpc to the South-East from
the nucleus and generated an expanding asymmetric
density wave in the disk of Arp~10. The position of maximum density perturbation 
was initially in the South-Eastern part of Arp~10 but later relocated to the North-Western part
due to rotation. The companion is now seen 3~kpc to the South-West of the nucleus.

\section{Numerical modeling of ARP~10}
\label{s7}

In this section we perform a simplified two-dimensional hydrodynamic modeling of ARP~10. We want 
to see if the collision with a likely intruder galaxy can reproduce the observed 
sizes of the inner and outer rings and expansion velocity 
of the outer ring.
We assume that the rings of ARP~10 are a manifestation 
of gaseous (rather than stellar) density waves that propagate in the gas disk 
of the galaxy. This conjecture is motivated by the fact that most disk galaxies
have their gas disks extending far beyond stellar disks. The spectral modeling
in \S~\ref{s5.2} suggests that the radial scale length of the stellar disk
in Arp~10 ($\approx 7.2$~kpc) is much smaller than that of the gas disk. 
A neutral hydrogen imaging of Arp~10 by
\citet {CA96} also shows that the gas disk extends at least 2.7 times the
mean size of the outer ring.
Therefore, we neglect density perturbations in the stellar component
of Arp~10 and assume that the stellar disk serves merely as a source of axisymmetric
gravitational potential. 

\subsection{Model description and basic equations}
\label{s7.1}

Our model galaxy consists of a thin self-gravitating gas disk,
which evolves in the combined gravitational potential of a
spherical dark matter halo, spherical bulge, thin stellar disk, and
companion galaxy. The gas disk is assumed to
be isothermal at $T=9000$~K.  We use the thin-disk approximation to
describe the motion of gas in Arp~10. In this approximation, the
radial extent of the gas disk is assumed to be much larger than its
vertical height, discarding the need to solve for the vertical
motion of gas. A near-vertical impact may send ripples through the
gas disk of a target galaxy. We assume that these vertical
oscillations have little effect on the appearance of Arp~10 due to
its low inclination.  

The basic equations governing the dynamics of the gas disk in Arp~10 are
\begin{equation}
{\partial \Sigma_{\rm g} \over \partial t} = - {\bl \nabla} \cdot ({\bl v} \Sigma_{\rm g}),
\label{first}
\end{equation}

\begin{equation}
{\partial {\bl v} \over \partial t} +({\bl v} \cdot {\bl
\nabla}) {\bl v} = -{\bl \nabla \Phi_{\rm g}}  -{\bl \nabla} \Phi_{\rm s,b,h,i} - {{\bl \nabla} P
 \over \Sigma_{\rm g}}.
 \label{second}
\end{equation}
Here, $\Sigma_{\rm g}$ is the gas surface density, $\bl v$ is the gas velocity
in the disk plane, $P=c_{s}^2 \Sigma_{\rm g}$ is the vertically integrated gas pressure, and
$c_{\rm s}$ is the sound speed. The gravitational potentials of the gas disk $\Phi_{\rm g}$,
bulge $\Phi_{\rm b}$, stellar disk $\Phi_{\rm s}$, dark matter halo $\Phi_{\rm h}$, and 
companion $\Phi_{\rm i}$ are provided below.

In order to compute $\Phi_{\rm g}$, $\Phi_{\rm b}$, $\Phi_{\rm s}$, and $\Phi_{\rm h}$, 
we have to make assumptions
about the total masses and radial profiles of gas, stars, and dark matter in Arp~10.
The total atomic hydrogen mass in Arp~10 is $1.1\times 10^{10}~M_\odot$ 
\citep[][corrected for a distance of 130 Mpc]{CA96}. Hence, we set the total gas mass (atomic hydrogen plus helium)
within a 50~kpc radius to be $1.4\times 10^{10}~M_\odot$. We neglect a possible
contribution of molecular hydrogen because its mass is highly uncertain.
The radial distribution of gas in the pre-collision Arp 10 is difficult to constrain. 
We assumed that the gas distribution in the pre-collision ARP~10 
exponentially declined with radius and tried different 
values for the central gas surface density $\Sigma_{\rm g0}$ and exponential scale length 
$r_{\rm g0}$. We found that the Toomre $Q$-parameter is larger or equal 2.0 throughout the disk
irrespective of $\Sigma_{\rm g0}$ and $r_{\rm g0}$.
The $Q$-parameter may decrease below 2.0 only if
$r_{\rm g0}$ is much smaller than the exponential scale length of the stellar 
disk $h_{\rm old}=7.2$~kpc and the gas disk in the pre-collision Arp~10
was strongly concentrated to its center. The latter is unlikely because gas disks usually 
extend further in radius than stellar disks \citep[see ][ for the
case of Arp~10]{CA96} and thus have shallower radial profiles.
This implies that the pre-collision Arp~10
might have depleted its gas content and reached a tentative state of gravitational stability
\citep[see e.g.][]{Zasov96}. We note that high values of the $Q$-parameter are typical for 
gas disks in low surface brightness galaxies.
For our simulations we adopt $r_{\rm g0} = 12$~kpc  and $\Sigma_{\rm g0}=22~M_\odot$~pc$^{2}$. 
We find that these values reproduce best both the radial and azimuthal distributions of H$\alpha$ 
surface brightness in Arp~10. 
The gravitational potential of a thin gas disk is then calculated by (Binney \& Tremaine, 
Sect.\ 2.8) 
\begin{eqnarray} 
  \Phi_{\rm g}(r,\phi) &=& - G \int_0^\infty r^\prime dr^\prime \times \nonumber \\
               &\times&  \int_0^{2\pi} 
               \frac{\Sigma(r^\prime,\phi^\prime) d\phi^\prime} 
                    {\sqrt{{r^\prime}^2 + r^2 - 2 r r^\prime 
                       \cos(\phi^\prime - \phi) }}. 
\end{eqnarray} 
This sum is calculated using a FFT technique which applies the 2D Fourier 
convolution theorem for polar coordinates.

The present-day rotation curve of Arp~10 is used to set constrains on the total masses of
stellar and dark matter components. 
The Arp~10 rotation curve flattens at a radius of approximately 3~kpc 
and slowly rises at larger radii. We use a steady-state equation of motion for the azimuthal component 
of the gas velocity to fit the model and
measured rotation curves. We assume that the thin stellar disk of Arp~10 has an exponentially 
declining surface density profile, whereas the spherical bulge of Arp~10 has an exponentially declining 
volume density profile.
The exponential scale lengths of the bulge $h_{\rm b}=1.77$~kpc and stellar disk 
$h_{\rm old}=7.2$~kpc are known from the R-band photometry of 
\S~\ref{s2}. 
For the dark matter distribution we adopt a modified isothermal sphere
\begin{equation}
\rho_{\rm h}={\rho_{\rm h0}\over (1+r/r_{\rm h})^2},
\end{equation}
where $\rho_{\rm h0}$ and $r_{\rm h}$  are the central volume density and characteristic
scale length of the dark matter halo. The central densities of the stellar disk $\Sigma^{\rm old}_{\rm 0}$ 
and the bulge $\rho_{\rm b0}$, and the dark matter halo profile are free parameters.

We find that the stellar disk and bulge (if considered alone) produce a declining rotation 
curve at radii larger than 10~kpc
and hence cannot account for the shape of the rotation curve in Arp~10, irrespective of the 
total stellar mass. This means that Arp~10 must possess a massive dark
matter halo. The exact mass 
partition between different components is somewhat uncertain. In fact, a strongly concentrated 
dark matter halo can approximately reproduce the Arp~10 rotation curve without invoking 
the need for the stellar disk or bulge. However, it would contradict the
presence of a considerable old stellar population in Arp~10 as implied
by spectral modeling in \S~\ref{s5.2}.
After experimenting with different values of the free parameters, we find that the  
rotation velocity of Arp~10 at large radii ($\approx 320$~km~s$^{-1}$) is best reproduced by a massive halo with
$M_{\rm h}=8\times 10^{11}~M_\odot$, $r_{\rm h}=1.5$~kpc, and $\rho_{\rm h0}=0.6~M_\odot$~pc$^{-3}$.
We also include a stellar disk and bulge with masses and central densities
$M_{\rm s}=1.0\times 10^{11}~M_\odot$, $\Sigma^{\rm old}_{\rm 0}=300~M_\odot$~pc$^{-2}$ and 
$M_{\rm b}=4\times 10^{10}~M_\odot$, $\rho_{\rm b0}=0.3~M_\odot$~pc$^{-3}$, respectively.
All masses are calculated within a 50~kpc radius of Arp~10.
The masses of the stellar disk and bulge are in fact the upper limits, because a more massive 
stellar disk and bulge would generate a declining rotation curve at radii larger than 15 kpc,
which apparently contradicts the slowly rising rotation curve of Arp~10.
We note that the parameters of the stellar disk \citep[see][]{Bizyaev03,Bizyaev04}, 
when taken together with high values of the $Q$-parameter,  strongly suggest a low 
surface brightness nature of the pre-collision Arp~10.
It is not unexpected that interacting (colliding) galaxies may have had
low surface brightness progenitors. This is because interactions usually produce a
new specific structure, which is much easier to see if the progenitor galaxies
have been of low surface brightness. Otherwise, the new structure may wash
out unless the interaction is very strong. The detection of a low surface
brightness progenitor of the interacting galaxy Arp~82 \citep{Hancock07} adds more
credit to this hypothesis.

\begin{figure}
\plotone{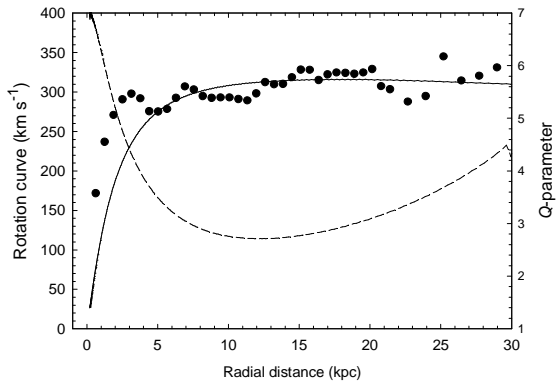}
\caption{Model rotation curve (solid line) that is assumed in
hydrodynamic simulations, radial distribution of the 
Toomre $Q$-parameter (dashed line), and observed rotation
curve of Arp~10 (filled circles).}
\label{f12a}
\end{figure}

The model rotation curve (solid line), radial distribution of 
the $Q$-parameter (dashed line) in the pre-collision Arp~10, and the 
measured rotation curve (filled circles) of the present-day Arp~10 
are shown in Figure~\ref{f12a}.
The model rotation curve shows good agreement with the measured curve except for
the inner 3~kpc, where the measured curve rises steeper than the model one. 
We assume that the steep rise is probably
caused by material that flows into the inner ring from the
inter-ring region. Such a radial inflow of gas can indeed be seen
in the model velocity field obtained in the next section. 
Hence, the steep rise of the measured rotation
curve could be a manifestation of angular momentum
conservation and the pre-collision Arp~10 might have had 
no peak at 3~kpc and shallower rotation curve in the central region.

Once the parameters of the stellar disk, bulge, and the dark matter halo are determined, we
can calculate their gravitational potentials. The gravitational potential of a thin stellar disk
in the plane of the disk, the surface density of which declines exponentially with radius, 
can be evaluated as \citep{BT}
\begin{equation}
\Phi_{\rm s}=-\pi G \Sigma^{\rm old}_{\rm 0} r \left[ I_0(y) K_1(y) -I_1(y) K_0(y) \right], 
\end{equation}
where $y=r/(2h_{\rm old})$ and $I_{\rm n}$ and $K_{\rm n}$ are the modified Bessel functions of the first 
and second kinds. The radial gravity force per unit mass of a spherical bulge 
is evaluated as 
\begin{equation}
{\partial \Phi_{\rm b} \over \partial r} = {G M_{\rm b}(r)\over r^2},
\end{equation}
where $M_{\rm b}(r)$ is the mass of the bulge contained within radius $r$.
The radial gravity force per unit mass of a spherical dark matter 
can be written as 
\begin{equation}
{\partial \Phi_{\rm h} \over \partial r}=4 \pi G \rho_{\rm h0}
r_{\rm h}\left[ r/r_{\rm h} - \arctan(r/r_{\rm h}) \right]
\left({r_{\rm h}\over r}\right)^2.
\label{halo}
\end{equation}
The companion is modeled as a point object with 
mass $M_{\rm c}=2\times 10^{11}~M_\odot$ (see \S~\ref{s6}),
which moves in the combined gravitational potential of the 
dark matter halo and bulge.
The equation of motion for the companion is solved using the
forth-order Runge-Kutta method with adaptive step size control.  The
gravitational potential of the companion $\Phi_{\rm i}$ is
calculated using a softened point-mass potential as described in
\cite{BT}, sect. 2.8. The softening parameter is set to 5~kpc.

\subsection{Code description}
\label{s7.2}

An Eulerian finite-difference code is used to solve equations~(\ref{first})-(\ref{second})
in polar coordinates ($r,\phi$). The basic algorithm of the code is similar
to that of the ZEUS code presented by \cite{SN}. The operator
splitting is utilized to advance in time the dependent variables in two
coordinate directions. The advection is treated using the consistent
transport method of Stone and Norman and the van Leer interpolation  scheme.
The timestep is determined according to the usual Courant-Friedrichs-Lewy criterion.
The numerical grid has $512\times 512$ points, which are uniformly spaced in 
the azimuthal direction and logarithmically spaced 
in the radial direction, i.e., the radial grid points are chosen to lie between the inner
boundary $r_{\rm in}=0.2$~kpc and outer boundary $r_{\rm out}=50$~kpc so that 
they are uniformly spaced in $\log r$. 

\subsection{Results of numerical modeling}
\label{s7.3}

\begin{figure}
\plotone{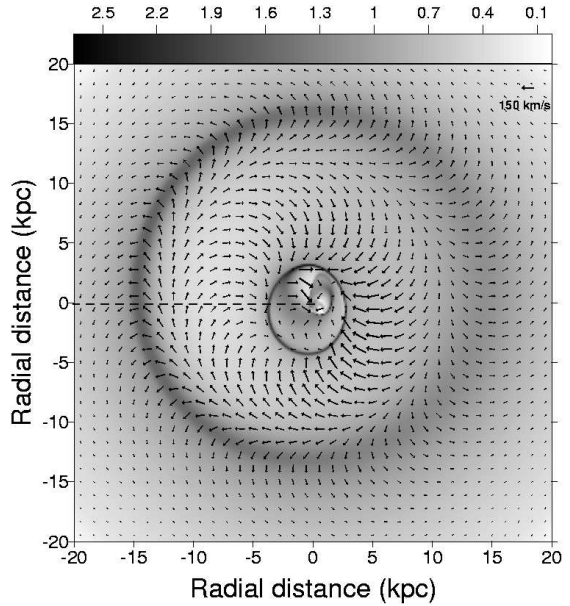}
\caption{Model residual velocity field superimposed on the gas surface
density distribution at $t=85$~Myr after the collision.
The velocity of $150~km~s^{-1}$ is indicated in the upper-right corner
and scale bar on the top is in $M_\odot$~pc$^{-2}$. 
The dashed line shows the radial cut along which the circular 
velocity of gas is determined.}
\label{f12}
\end{figure}

We model the collision by releasing the companion at 50~kpc above the 
disk plane of Arp~10. The initial velocity of the companion is 490~km~s$^{-1}$. 
It falls near axially and hits Arp~10 at a distance of approximately 
2.7~kpc from the center. At the time of impact, the companion velocity relative to Arp~10 is 
approximately 800~km~s$^{-1}$. The gravitational perturbation from the companion 
sets off expanding density waves in the gas disk of Arp~10. 
Figure~\ref{f12} shows the residual velocity 
field superimposed on the map of the gas surface density in Arp~10 at $t\approx 85$~Myr
after the collision. The residual velocity field is obtained by subtracting
the circular velocities from the proper velocities of gas.
Since we were limited by the two-dimensional approach, we could not fine-tune
the numerical model to reproduce the exact visual appearance of Arp~10.
Nevertheless, we were able to match the mean radii of the inner and outer rings,
$\approx 3$~kpc and $\approx 14$~kpc, respectively. 
An overall expansion of the outer ring is clearly seen. The model radial expansion velocity
of the outer ring as a function of azimuth at $t=85$~Myr is shown by the solid line 
in Figure~\ref{f5.2}. The maximum expansion velocity is approximately 120~km~s$^{-1}$, 
which is in good agreement with a measured maximum expansion velocity of
$110 \pm 10$~km~s$^{-1}$.

The flow of gas in the vicinity of the
inner ring is complicated. The gas appears to stream toward the inner ring from 
the inter-ring region of Arp~10. This inward streaming motion of gas provides an extra compression
to the inner ring, the gas surface density of which becomes an order of magnitude higher than
that of the outer ring. 

\begin{figure}
\plotone{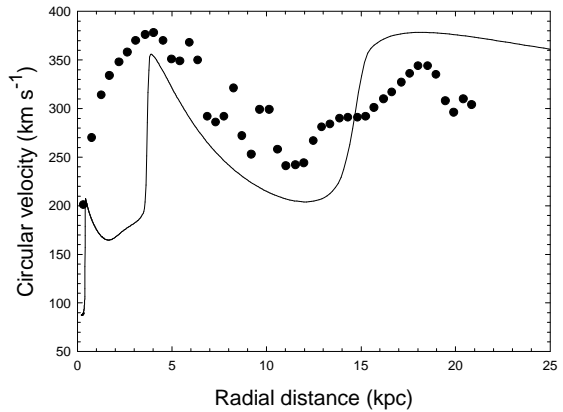}
\caption{Comparison of radial distribution of the model circular
velocities obtained along the radial cut in Figure~\ref{f12} (solid line)
with radial distribution of observed circular velocities along the upper 
half of radial cut~2 in Figure~\ref{f2} 
(filled circles).}
\label{fvelcut}   
\end{figure}

If outwardly propagating gas density waves are {\it axisymmetric}, then they can be 
identified by characteristic signatures in the azimuthal flow of gas in a galactic disk. 
Indeed, as the wave propagates outward, it pools the matter from both sides of the wave toward its
current location.  This should alter the circular velocity of gas $V_\phi$ on both sides of the wave
due to conservation of angular momentum. More specifically, the angular velocity of gas 
that lies ahead of the wave should increase, whereas the angular velocity of gas that lies 
behind the wave should decrease.  Of course, any considerable deviation from
axial symmetry of the wave would break the conservation of angular momentum.
Nevertheless, these characteristic signatures can be detected 
in the radial distribution of $V_\phi$, if the latter is constructed along a narrow 
radial cut rather than being azimuthally averaged. Filled circles in Figure~\ref{fvelcut} show
the radial distribution of $V_\phi$ obtained along the upper half of cut~2 in Fig~\ref{f2}. 
A characteristic double-peaked shape is conspicuous. The outer and inner rings 
are located at approximately 15~kpc and 3~kpc, respectively. It is seen that
$V_\phi$ is growing across the outer ring from 11~kpc to 18~kpc, exactly as predicted
from conservation of angular momentum of gas that is pooled toward the ring. 
The radial profile of $V_\phi$ has a local minimum at
11~kpc, a radius at which the flow of gas is reversed from expansion to contraction 
toward the inner ring (see Figure~\ref{f12}). As a result, $V_\phi$ starts to grow again at smaller radii 
and reaches a maximum near the position of the inner ring. Inside the inner ring, $V_\phi$
quickly drops to a minimum. Similar double-peak radial profiles of $V_\phi$ were found along the 
radial cuts taken at $\pm 10^\circ$ and $\pm 20^\circ$ away from cut~2. We note that the rotation curve in Figure~\ref{f5}
does not have a well-defined double-peaked shape, probably due to azimuthal averaging along circular rings.
We compare the measured radial profile of $V_\phi$ with that obtained in our numerical modeling.
That is, we take a radial cut through the densest region in the outer ring (dashed line
in Figure~\ref{f12}) and plot the resulting
radial profile of $V_\phi$ along this cut by the solid line in Figure~\ref{f12}.
Although a quantitative agreement is difficult to attain in our simplified thin-disk
model due to neglected vertical dynamics of Arp~10,  
a qualitative agreement is clearly seen and is encouraging. We consider
a characteristic double-peak radial profile of circular velocity
to be powerful evidence in favor of the collisional origin of Arp~10.

A good agreement between the modeled and measured sizes and maximum expansion velocities
of the rings in Arp~10 suggests that approximately 85~Myr have passed since the collision.
Our simulations indicate that the velocity of the companion has dropped to approximately 
480~km~s$^{-1}$ at $t=85$~Myr after the collision,
which is also in agreement with the measured velocity of the companion.
The first resonant ring forms soon after the collision so that the age of the outer
ring should be equal or less than 85~Myr. The spectroscopic analysis in \S~\ref{s5} 
suggests a similar age, 90~Myr. 

The distribution of H$\alpha$ emission from the outer ring of Arp~10 is characterized by
a crescent shape (Figure~\ref{f2}). We test a conclusion of \citet{CAM} that this peculiar H$\alpha$ morphology 
is a result of both the off-center collision and star formation threshold in the gas 
disk of Arp~10. We model the H$\alpha$ surface brightness in Arp~10 using a method
described in \cite{Vorob03}. The star formation rate is assumed to obey 
a Schmidt law \citep{Kennicutt98} in the form 
\begin{equation}
\Sigma_{\rm SFR}=2.5\times 10^{-4}\, \Sigma_{\rm g}^{1.5},
\label{Schmidt}
\end{equation}
where the star formation rate $\Sigma_{\rm SFR}$ and gas surface density $\Sigma_{\rm gas}$
are measured in units of $M_\odot$~yr$^{-1}$~kpc$^{-2}$ and $M_\odot$~pc$^{-2}$, respectively.
We modify the Schmidt law~(\ref{Schmidt}) by assuming that star formation terminates if
the Toomre $Q$-parameter exceeds $Q_{\rm cr}=1.6$, a critical value for stability of gas disks 
(characterized by a flat rotation curve) against local non-axisymmetric perturbations 
\citep{Polyachenko}. The star formation rate is converted to the H$\alpha$ luminosity 
using a calibration of
\cite{Kennicutt83} for solar metallicity
\begin{equation}
SFR~(M_{\odot}~{\rm yr^{-1}})~=~7.9\times10^{-42}~L_{\rm H\alpha}~({\rm
erg~s^{-1}}).
\label{halpha}
\end{equation}
Metal-poor stars produce more ionizing photons than metal-rich stars.
The star-forming regions in Arp~10 have lower than 
the solar metallicity and
equation~(\ref{halpha}) is expected to give a lower limit on the H$\alpha$ luminosity.
The extinction in the  gas disk of Arp~10 is taken into account 
by estimating  $A_{\rm v}$ from the model's known gas surface density
via the relation given in \citet{Bohlin78} for $R_{\rm v}=3.1$.
The model H$\alpha$ fluxes are then corrected for the internal 
extinction $A_{\rm H\alpha}=0.75\: A_V$ \citep{Cardelli}.

\begin{figure}
\plotone{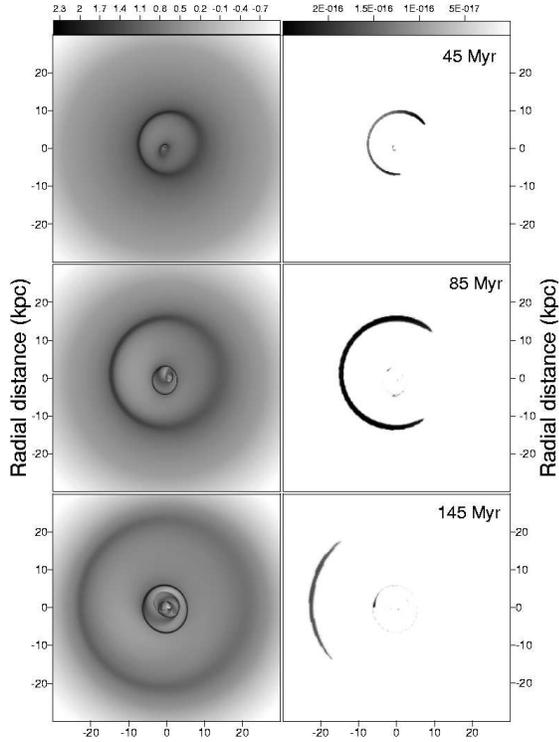}
\caption{Model gas surface density (left) and H$\alpha$
surface brightness (right) distributions at three consecutive times after
the collision: 45~Myr, 85~Myr, and 145~Myr. The scale bars are 
in $M_{\odot}$~pc$^{-2}$ (left) and 
erg~s$^{-1}$~cm$^{-2}$~arcsec$^{-2}$ (right)}.
\label{f13}
\end{figure}

Figure~\ref{f13} shows the gas surface density distribution (left)
and H$\alpha$ surface brightness distribution (right) at three consecutive times
after the collision $t=45$~Myr, 85~Myr, and 105~Myr. 
The azimuthal distribution of H$\alpha$ surface brightness in the outer ring has 
a characteristic crescent shape. The values of model H$\alpha$ surface brightness
$\Sigma_{\rm H\alpha}$ in the outer ring agree approximately with the values measured by \citet{CAM}.
More specifically, the measured variation of $\Sigma_{\rm H\alpha}$ 
is between $1.1\times 10^{-16}$~erg~s$^{-1}$~cm$^{-2}$~arcsec$^{-2}$ for the faintest part
of the outer ring and $6.7\times10^{-16}$~erg~s$^{-1}$~cm$^{-2}$~arcsec$^{-2}$ 
for its brightest part. These values are not corrected for the internal
extinction.  The model predicts values equal or less than 
$2\times10^{-16}$~erg~s$^{-1}$~cm$^{-2}$~arcsec$^{-2}$
but corrected for the mean internal extinction in the outer ring 
$A_{\rm H\alpha}\approx 1.3$~mag as described above.

Comparison between H$\alpha$ and gas distributions
clearly indicates that the H$\alpha$ emission originates from regions with the highest
gas surface density. These regions are characterized by $Q < Q_{\rm cr}=1.6$, while
the rest of the outer ring has $Q>Q_{\rm cr}=1.6$. Significant azimuthal variations 
of the $Q$-parameter in the outer ring are easiest to reproduce
by a slightly off-center  collision (2.7~kpc), while
a bull-eye central collision would require for more sophisticated physical conditions
in the outer ring (e.g. systematic variations in gas temperatures along the ring perimeter).
Thus we confirm that an off-center collision and star formation threshold are most probably 
responsible for the lack of H$\alpha$ emission from the North-Eastern part of the 
Arp~10 outer ring, a conclusion previously made by \citet{CAM}. 
Our modeling predicts that the inner ring should be significantly less prominent in H$\alpha$
than the outer ring.  This seems to contradict the observations. We attribute this disagreement
to the adopted gas isothermality in our numerical model -- high gas densities in the inner ring
might have invoked  substantial cooling, which had lowered the gas temperature and 
associated star formation threshold.

It is interesting to consider the model-predicted temporal
variations in the H$\alpha$ surface brightness shown in
Figure~\ref{f13} (right frames). In the early phase of evolution,
Arp~10 was less prominent in H$\alpha$, mainly due to a higher gas
density and associated extinction in the expanding rings. At the
present epoch, Arp~10 is likely to have reached a peak in the
H$\alpha$ surface brightness. In the next 20~Myr, Arp~10 is
expected to fade in H$\alpha$, mainly due to a drop in the gas
surface density of the outer ring. 
The radius of the stellar disk in Arp~10 at the $\mu_B=25$~mag~arcsec$^{-2}$ isophote
is $D_{25}/2$ = 28~kpc, according to RC3 \citep{RC3}.
The middle-right panel in Figure~\ref{f13}
indicates that the present H$\alpha$ ring in Arp~10 is located at about
half the radius of the stellar disk. This tendency was also found by
\citet{Romano07} for a sample of 15 northern ring galaxies. Thus we argue
that a preferable location of H$\alpha$ rings at half the stellar disk
radius is probably caused by high extinction in the early expansion phase (which
obscures H$\alpha$ emission and hinders detection of small rings) and by a drop
in the gas surface density below the star formation threshold in the later
expansion phase. We caution that in the real systems there may 
be other mechanisms
such as an interaction between rings and spiral density waves, which affect local star 
formation and H$\alpha$ fluxes. We also note that the observed
outer ring in Arp~10 has a noticeably larger eccentricity than the model
outer ring. This is not a geometry effect because of low inclination of Arp~10 
($i=23^\circ$).  We attribute a low eccentricity of our model outer ring 
to several factors. These are the two-dimensional nature of our modeling,
initially axisymmetric distribution of gas in the disk of pre-collision Arp~10,
and use of a fixed dark matter halo that cannot move in responce 
to the companion.

\subsection{Global star formation rate}
\label{s7.4}

The gas-to-stellar mass ratio in Arp~10 is about $9 \%$
within  a 50~kpc radius, which is typical for spiral galaxies. The global star formation
rate (SFR) in Arp~10 can be estimated either from far-infrared (FIR) or
H$\alpha$ fluxes. The H$\alpha$ flux is affected by the
internal extinction, which according to \S~\ref{s4.1} 
may vary considerably across the disk of Arp~10. 
Therefore, we use the FIR fluxes taken from the NED database to estimate the FIR luminosity
of Arp~10. The FIR fluxes of 0.784 Jy (60 $\mu$m) and 1.767 Jy (100 $\mu$m)
suggest the FIR luminosity $3.84\times 10^{10}~ L_{\odot}$. 
This value corresponds to the SFR of 21~$M_{\odot}$~yr$^{-1}$, if a conversion factor between the FIR luminosity and SFR 
from \citet{Thronson86} is used in combination with a \citet{Kennicutt83} initial mass function.  
We note that according to \citet{Zasov95} the conversion factor 
can be almost twice lower than the value derived by \citet{Thronson86}.
Hence, we set a lower limit on the SFR in Arp~10 to be 10 $M_{\odot}$~yr$^{-1}$.
This value is still a factor of two larger than that derived by \citet{CAM}
using the H$\alpha$ photometry. \citet{Romano07} also found that SFRs estimated
from FIR fluxes are also larger than those estimated from H$\alpha$ fluxes.
We attribute this difference to the internal extinction, which lowers H$\alpha$-estimated 
SFRs. 
The SFR in Arp~10 is very similar to that inferred by \citet{Mayya05} for
the Cartwheel.

On the other hand, our numerical simulations predict the present
SFR in Arp~10 to be about $6~M_\odot$~yr$^{-1}$. If our FIR-based
value of 10 -- 21 $M_\odot$~yr$^{-1}$ is correct, it means that
the star formation efficiency in Arp~10 is twice to four times more
efficient than that derived by \citet{Kennicutt98} and used by us
in equation~(\ref{Schmidt}). This elevated efficiency of star
formation appears to be a common feature of ring galaxies. For
instance, the Cartwheel galaxy is also characterized by a star
formation efficiency three times larger than that of isolated and
interacting galaxies in the Kennicutt's sample \citep{Vorobyov03}.

Our numerical simulations indicate that the SFR in Arp~10 was a
nearly linear function of time during the last 85~Myr. Hence, the
present epoch SFR of 10 -- 21 $M_{\odot}$~yr$^{-1}$ suggests 
that $(4 - 9) \times 10^{8} ~M_{\odot}$ 
of young stars were born as a result of the passage of the density wave.
This is less than $1\%$ of the total stellar disk mass.

At the same time, Arp~10 has converted 3 -- $6\%$ of the
total gas mass into stars. These values have to be certainly taken as
lower limits for the star formation efficiency because only approximately half of 
the total gas content is located in the inner 16~kpc.

\section{Limitations of our approach}
\label{s8}

The spectral and numerical modeling of Arp~10 performed by us has
several limitations and simplified assumptions, the impact of which
on our results we discuss below. An assumption of constant
propagation velocity of the rings made in \S~\ref{s4} contradicts
numerical simulations \citep[see e.g.][]{Vorobyov03}. In the early
evolutionary stage, the rings propagate faster than in the late
stage. This tendency can be seen in Figure~\ref{f5.2}, which shows
the azimuthal profiles of the gas expansion velocity in the outer
ring at t=85~Myr (solid line) and t=105~Myr (dashed line) after the
collision. The assumption of constant propagation velocity introduces
additional uncertainty to the spectral modeling.

A faint signal and low-quality of long-slit spectra in the inter-ring
region and behind the outer ring introduces an uncertainty in the
spectral index modeling. Because of the adopted thin-disk
approximation, we cannot model the vertical motions in the disk of
Arp~10 after the collision, although these motions are well seen in
the velocity fields in the inner parts of the galaxy. Ideally, a
three-dimensional hydrodynamic model of Arp~10 should be combined
with a population synthesis model, and the resulting model colors and
spectra should be compared with observational data. 
High-resolution maps of gas density and internal extinction of Arp~10
would be needed in this case.

Nevertheless, the observed kinematics, spectral index modeling, and
hydrodynamic modeling of Arp~10 are in general agreement.  This
suggests that we envision correctly a general picture of the
collision and subsequent star formation in Arp~10. We postpone a more
detailed numerical and spectral modeling to a future paper.

\section{Conclusions}
\label{s9}

We apply kinematic, spectral, and numerical approaches to 
develop a self-consistent model of propagating star formation in a 
collisional ring galaxy Arp~10. Our kinematic data have
the best spatial resolution obtained for Arp~10 to date.

We found that the pre-collision Arp~10 was a large (Sb?) disk
galaxy with an extended low surface density exponential stellar
disk (scale length 7.2 kpc), substantial exponential bulge, and
massive dark matter halo.  The spiral arms that are traceable in
our B-band and H$\alpha$ images beyond the outer ring suggest that
Arp~10 has had a spiral structure before the collision. Our
spectral index modeling suggests the solar (using all selected
indices or Mg, Ca, and Ti-related indices) or slightly subsolar
(using iron-dominated spectral indices) metallicity of the old
stellar population in the center of Arp~10. The gradient of
relative metallicity of the old population is
$-0.08$~dex~kpc$^{-1}$.

The available measurements of the total hydrogen mass and the shape
of the rotation curve suggest high values of the Toomre
$Q$-parameter ($>2.6$) in the pre-collision gas disk of Arp~10,
which implies that star formation was moderated or even totally
suppressed before the collision. The $Q$-parameter may decrease
below a tentative stability limit of 1.6 if either a substantial
undetected amount of molecular gas is present or the gas disk is
highly concentrated to the center. Both possibilities appear to be
unlikely.  From the spectra of Arp~10 we estimate the oxygen
abundance to be 12+log(O/H)=8.6 with little variation along the
radius.

Some 85 Myr ago an early-type spiral galaxy (Sab?) passed slightly 
off-center and near-axially through the disk of Arp~10. The velocity 
of the intruder galaxy was about 800~km~s$^{-1}$ at the time of closest 
encounter and its mass was about $1/4$ the mass of the target galaxy. 
The intruder is now seen as a small "knot" 5 arcsec 
to the South-West from the nucleus of Arp~10.
The gravitational perturbation from the intruder generated two outwardly
expanding density waves, which are presently seen in the B-band 
and H$\alpha$ images as bright asymmetric rings. 
The expansion 
velocity of gas in the outer ring is a strong function of azimuth, 
with a maximum value of about $110 \pm 10$~km~s$^{-1}$ 
in the brightest part of the ring and
a minimum value of about $25 \pm 5$~km~s$^{-1}$.
The inner ring has a more symmetric 
shape than the outer ring and expands with a velocity less than
50~km~s$^{-1}$. The propagating density waves in Arp~10 do not 
generate strong shock waves.

Star formation is triggered at the crest of an expanding gas density
wave, if the gas surface density exceeds a star formation threshold.
An apparent crescent-like distribution of H$\alpha$ emission in the
outer ring is most probably caused by a variation of the Toomre
$Q$-parameter along the ring.  The current star formation rate is
estimated from the far-infrared flux to be 10 --
21 $M_{\odot}$~yr$^{-1}$. The expanding waves of star formation have
created less than $1\%$ of the total stellar disk mass since the
collision.


\acknowledgements

This project was partially supported by grants RFBR 04-02-16518,
RFBR 05-02-16454, RFBR 06-02-16819-a, and Federal Agency of Education 
(project code RNP 2.1.1.3483). We are grateful to the anonymous 
referee for very useful comments and suggestions.
This research is based on observations collected 
with the 6-m telescope of the Special Astrophysical Observatory (SAO)
of the Russian Academy of Sciences (RAS), operated under the
financial support of the Science Department of Russia
(registration number 01-43.)
This research has made use of the NASA/IPAC
Extragalactic Database (NED) which is operated by the Jet
Propulsion Laboratory, California Institute of Technology, under
contract with the National Aeronautics and Space Administration. We
acknowledge the usage of the HyperLeda database
(http://leda.univ-lyon1.fr). 
E.I.V. acknowledges support from a CITA National Fellowship. 
Numerical simulations were done on
the Shared Hierarchical Academic Research Computing Network (SHARCNET).


\clearpage

\begin{table}
\begin{center}
\caption{Emission lines in Arp~10 \label{t2}}
\begin{tabular}{rccccccc}
\tableline\tableline
$r$, kpc& \multicolumn{7}{c}{Fluxes in lines, in units of $10^{-16}$~erg~s$^{-1}$~cm$^{-2}$ }\\
 & H$\beta$ & H$\alpha$ & [NII]6548 & [NII]6584 & [OII]3728 & [OIII]5007 &
[OIII]4959 \\
\tableline
-1.36 & 8.52$\pm$0.34& 27.6$\pm$0.32& 2.93$\pm$0.12& 9.66$\pm$0.15& 2.74$\pm$0.23& 0.48$\pm$0.12&    --    \\  
 1.36 & 8.06$\pm$0.35& 23.1$\pm$0.28& 4.43$\pm$0.18& 8.63$\pm$0.14& 2.36$\pm$0.26& 0.24$\pm$0.06&    --    \\
11.59 & 4.60$\pm$0.10& 11.3$\pm$0.14& 1.10$\pm$0.04& 4.10$\pm$0.07& 4.53$\pm$0.19& 0.85$\pm$0.04& 0.22$\pm$0.03\\
14.54 & 6.15$\pm$0.09& 21.4$\pm$0.23& 1.90$\pm$0.03& 7.98$\pm$0.10& 4.48$\pm$0.15& 2.51$\pm$0.05& 0.34$\pm$0.02\\
16.36 & 2.67$\pm$0.05& 17.2$\pm$0.19& 1.37$\pm$0.04& 6.70$\pm$0.09& 1.19$\pm$0.09& 1.23$\pm$0.04& 0.23$\pm$0.02\\
\tableline
\end{tabular}
\end{center}
\tablecomments{Emission line fluxes at different radii $r$ in Arp~10.}
\end{table}

\begin{table}
\begin{center}
\caption{Extinction and abundances in Arp~10 \label{t3}}
\begin{tabular}{rccccc}
\tableline\tableline
$r$ & $P$ & $A_V$ & 12+log(O/H) & f$_{cor}$(H$\alpha$) &
f$_{cor}$(H$\beta$) \\
kpc &  & mag  &  dex & $10^{-15}$~erg~s$^{-1}$~cm$^{-2}$ & 
erg~s$^{-1}$~cm$^{-2}$\\
\tableline
-1.36 & 0.132$\pm$0.018 & 0.64$\pm$0.03 & 8.57$\pm$0.22 & 4.63$\pm$0.34& 1.61$\pm$0.34 \\ 
 1.36 & 0.091$\pm$0.017 & 0.00$\pm$0.00 & 8.53$\pm$0.28 & 2.31$\pm$0.28& 0.81$\pm$0.35 \\ 
11.59 & 0.190$\pm$0.011 & 0.00$\pm$0.00 & 8.53$\pm$0.10 & 1.13$\pm$0.14& 0.46$\pm$0.10 \\ 
14.54 & 0.336$\pm$0.010 & 1.01$\pm$0.02 & 8.65$\pm$0.06 & 4.88$\pm$0.25& 1.70$\pm$0.09 \\ 
16.36 & 0.320$\pm$0.018 & 4.25$\pm$0.09 & 8.62$\pm$0.10 & 54.6$\pm$1.16& 19.0$\pm$0.39 \\ 
\tableline
\end{tabular}
\end{center}
\tablecomments{Extinction $A_V$ in HII regions, 
excitation parameter $P$ (see text), oxygen abundance 12+log(O/H),
and H$\alpha$ and H$\beta$ fluxes corrected for the extinction 
(f$_{cor}$(H$\alpha$), f$_{cor}$(H$\beta$)) in units 
of $10^{-15}$~erg~s$^{-1}$~cm$^{-2}$ at different radii $r$ in Arp~10.}
\end{table}

\end{document}